\documentclass{aa}
\usepackage{graphicx,natbib,pjournal}

\begin{document}
\title{The Cosmic Infrared
Background Resolved by {\it Spitzer}.}
\subtitle{Contributions of Mid-Infrared Galaxies
to the Far-Infrared Background.}

\author{H. Dole\inst{1}
	\and
        G. Lagache\inst{1}
	\and
	J.-L. Puget\inst{1}
        \and
	K. I. Caputi\inst{1}
	\and
	N. Fern\'andez-Conde\inst{1}
	\and
	E. Le Floc'h\inst{2,3}
	\and
	C. Papovich\inst{2}
	\and
	P.~G. P\'erez-Gonz\'alez\inst{2}
	\and
        G. H. Rieke\inst{2}
	\and
        M. Blaylock\inst{2}
  }

\authorrunning{H. Dole et al.}
\titlerunning{ The Cosmic Infrared
Background Resolved by {\it Spitzer}.}

\offprints{Herv\'e Dole \email{Herve.Dole@ias.u-psud.fr}}

\institute{Institut d'Astrophysique Spatiale (IAS), bat 121, F-91405 Orsay, France; 
	   Universit\'e Paris-Sud 11 and CNRS (UMR 8617)
       \and
	Steward Observatory, University of Arizona, 933 N Cherry Ave,
	Tucson, AZ, 85721, USA
	\and
	Associated with Observatoire de Paris, GEPI, 92195 Meudon, France
             }

\date{Received: 31 October 2005; Accepted: 18 February 2006.}

\abstract
{}
{We quantify the contributions of 24~$\mu$m galaxies to the
Far-Infrared (FIR) Background at 70 and 160~$\mu$m.  We provide new
estimates of the Cosmic Infrared Background (CIB), and compare it with
the Cosmic Optical Background (COB).}
{Using Spitzer data at 24, 70 and 160~$\mu$m in three deep fields, we
stacked more than 19000 MIPS 24~$\mu$m sources with $S_{24}
\ge 60 \mu$Jy at 70 and 160~$\mu$m, and measured the resulting
FIR flux densities.}
{This method allows a gain up to one order of magnitude in depth in
the FIR. We find that the Mid-Infrared (MIR) 24~$\mu$m selected
sources contribute to more than 70\% of the Cosmic Infrared Background
(CIB) at 70 and 160~$\mu$m. This is the first direct measurement of
the contribution of MIR-selected galaxies to the FIR CIB. Galaxies
contributing the most to the total CIB are thus $z \sim 1$ luminous
infrared galaxies, which have intermediate stellar masses. We estimate that
the CIB will be resolved at 0.9 mJy at 70 and 3 mJy at 160~$\mu$m.  By
combining the extrapolation of the 24~$\mu$m source counts below
analysis, we obtain lower limits of $7.1\pm 1.0$ and $13.4 \pm 
1.7$~nW~m$^{-2}$~sr$^{-1}$ for the CIB at 70
and 160~$\mu$m, respectively.}
{The MIPS surveys have resolved more than three quarters of the MIR
and FIR CIB. By carefully integrating the Extragalactic Background
Light (EBL) SED, we also find that the CIB has the same brightness as
the COB, around 24~nW~m$^{-2}$~sr$^{-1}$. The EBL is produced on
average by 115 infrared photons for one visible photon. Finally, the
galaxy formation and evolution processes emitted a brightness
equivalent to 5\% of the primordial electromagnetic background (CMB). }

\maketitle

%
\section{Introduction}
The Cosmic Infrared Background (CIB) is the relic emission at
wavelengths larger than a few microns of the formation and evolution
of the galaxies of all types, including Active Galactic Nuclei
(AGN) and star-forming systems
\cite[]{puget96,hauser98,lagache99,gispert2000,hauser2001,kashlinsky2005}.
Characterizing the statistical behavior of galaxies responsible for
the CIB -- such as the number counts, redshift distribution, mean
Spectral Energy Distribution (SED), luminosity function, clustering --
and their physical properties -- such as the roles of star-forming vs
accreting systems, the density of star formation, and the number of
very hot stars -- has thus been an important goal
\cite[]{partridge67}. The SED of the CIB peaks near 150~$\mu$m. It
accounts for roughly half of the total energy in the optical/infrared Extragalactic
Background Light (EBL) \cite[]{hauser2001}, although still with
some uncertainty \cite[]{wright2004,aharonian2005}. Since locally the
infrared output of galaxies is only a third of the optical one
\cite[]{soifer91}, there must have been a strong evolution of galaxy
properties towards enhanced Far-Infrared (FIR) output in the past.
Understanding this evolution requires interpretation of cosmological
surveys conducted not only in the infrared and submillimeter spectral
ranges, but also at other wavelengths \cite[]{lagache2005}.

The cryogenic infrared space missions IRAS (Infrared Astronomical
Satellite) and ISO (Infrared Space Observatory) provided us with
valuable insights to the IR-dominated galaxies in the Mid-Infrared
(MIR) and FIR \cite[for
reviews]{sanders96,genzel2000,dole2003a,elbaz2005,lagache2005}.  ISO
MIR surveys were able to resolve a significant fraction of the
15~$\mu$m CIB \cite[counts close to convergence]{elbaz99}. Using model
SEDs of galaxies \cite[for instance]{chary2001,xu2001,lagache2003},
the contribution of MIR-selected galaxies to the peak of the CIB
(around 140 to 170~$\mu$m) can be inferred. \cite{elbaz2002} derived
that $64\pm 38$\% ($16 \pm 5$ over $25 \pm 7$ nW m$^{-2}$ sr$^{-1}$)
of the 140~$\mu$m background is due to ISOCAM 15~$\mu$m galaxies,
whose median redshift is $z \sim 0.8$.

%
\begin{figure}[!t]
   \centering
   \includegraphics[width=0.5\textwidth]{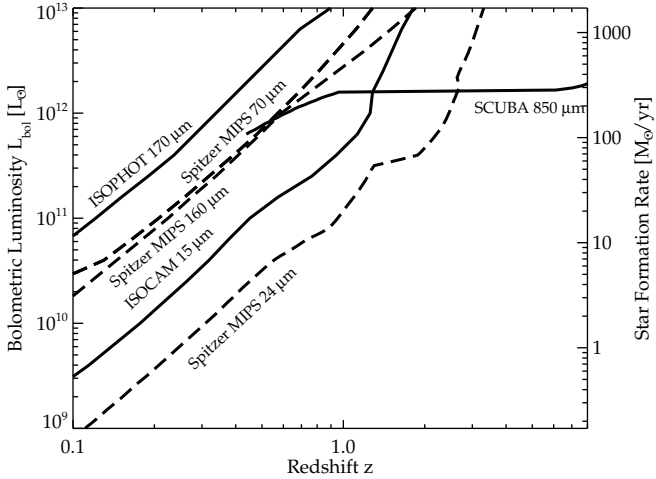}
   \caption{Sensitivity to the bolometric luminosity (and star
   formation rate, assuming star forming galaxies) of various infrared
   and submillimeter experiments. Detections of at least 10 sources in
   the surveys can be made in the areas above the curves.  We assumed
   the scenario of a typical deep survey.  {\it ISOCAM} 15~$\mu m$
   ($S_{\nu} > 250 \mu$Jy, 2 Sq. Deg.); {\it ISOPHOT} 170~$\mu m$
   ($S_{\nu} > 180 $~mJy, 5 Sq. Deg.); {\it Spitzer/MIPS} 24~$\mu m$
   ($S_{\nu} > 80 \mu$Jy, 5 Sq. Deg.); {\it Spitzer/MIPS} 70~$\mu m$
   ($S_{\nu} > 25 $~mJy, 5 Sq. Deg.); {\it Spitzer/MIPS} 160~$\mu m$
   ($S_{\nu} > 50 $~mJy, 5 Sq. Deg.); {\it SCUBA} 850~$\mu m$
   ($S_{\nu} > 1 $~mJy, 1 Sq. Deg.); This plot makes use of the
   Lagache et al. (2004) model and their starburst SED for the
   conversion to $L_{bol}$.  At $z \sim 1$, MIPS detects only ULIRGs
   in the FIR, and detects LIRGs in the MIR. The stacking analysis
   allows to gain an order of magnitude and to probe LIRGs in the
   FIR. } \label{fig:plot_lbol}
\end{figure}

The {\it Spitzer} Observatory \cite[]{werner2004} is performing much
deeper and wider-area surveys, in particular at 24, 70 to 160~$\mu$m
using the Multiband Imaging Photometer for {\it Spitzer} (MIPS)
\cite[]{rieke2004}. However, because of the limited angular resolution
(``smoothing'' the high spatial frequency signal in the FIR maps),
deep MIPS 70 and 160~$\mu$m maps are confusion limited
\cite[]{dole2003,dole2004b} -- the source surface density corresponds
to 20 beams per source or less e.g. in the GTO fields. The FIR images
do not allow us to directly probe the same galaxy population as that
detected at 24~$\mu$m, where the extragalactic source confusion is
less important.  Figure~\ref{fig:plot_lbol} shows the typical
sensitivity of MIPS surveys to the bolometric luminosity of galaxies
as a function of redshift, using the modeled starburst SED of
\cite{lagache2004}. At a redshift $z \sim 1$, MIPS FIR surveys are
sensitive to ultraluminous IR galaxies (ULIRGs, $L \ge 10^{12}
L_{\odot}$) where MIPS 24~$\mu$m surveys can probe luminous IR
galaxies (LIRGs, $L \ge 10^{11} L_{\odot}$).  It is therefor
impossible to derive MIR and FIR SEDs of individual LIRGs at $z
\sim 1$ and above.

MIPS can detect high redshift sources at 24~$\mu$m: about 25 to 30\%
of the population of galaxies lie at $z \ge 1.5$, at faint flux
densities (down to few tens of $\mu$Jy)
\cite[]{lefloch2004,egami2004,lonsdale2004,chary2004,houck2005,perez-gonzalez2005,caputi2005a}. 
\cite{papovich2004} showed that MIPS surveys resolve about 70\% of the
24~$\mu$m IR galaxy CIB for $S_{24} \ge 60~\mu$Jy. In comparison, MIPS
70 and 160~$\mu$m \cite[]{dole2004a} surveys detect only about 20\%
and less than 10\% of the CIB at 70 and 160~$\mu$m, respectively.
Programs of very deep 70~$\mu$m imaging on small fields
(e.g. D. Frayer, private communication) are likely to resolve a larger
fraction, but due to confusion noise the FIR CIB will still not be
significantly resolved into individual sources, while the MIR CIB at
24~$\mu$m is well resolved.

In this paper, we use a stacking analysis method that takes advantage
of the good sensitivity of the MIPS 24~$\mu$m MIR channel, to fill the
sensitivity gap between the MIR and the FIR surveys. By stacking the
FIR data at the locations of MIR sources, we statistically investigate
the FIR properties of 24~$\mu$m-selected galaxies. In particular, we
quantify the contribution of the 24~$\mu$m resolved galaxies to the 70
and 160~$\mu$m background, put strong lower limits to the CIB, and
give new estimates of the 70 and 160~$\mu$m background.

Throughout this paper, we adopt a cosmology with $h=0.65$, $\Omega_M =
0.3$ and $\Omega_{\Lambda} = 0.7$. The surface brightnesses (e.g. of
the CIB) are usually expressed in units of MJy/sr or nW m$^{-2}$
sr$^{-1}$. For a given frequency $\nu$ in GHz and wavelength
$\lambda$ in microns, the conversion between the two
is given by:
\begin{equation}
1 nW \, m^{-2} \, sr^{-1} = \frac{100}{\nu / GHz} MJy/sr =
\frac{\lambda / \mu m}{3000} MJy/sr
\end{equation}

%
\section{Data and Sample}
\label{section:DataSample}
The data for our analysis are from the {\it Spitzer} MIPS Guaranteed
Time Observations (GTO) cosmological surveys performed in three
fields: the Chandra Deep Field South (CDFS), the Hubble Deep Field
North (HDFN) and the Lockman Hole (LH). The MIPS observations at
24~$\mu$m are detailed in \cite{papovich2004} and at 70 and 160~$\mu$m
in \cite{dole2004a}. Each field covers about 0.4 square degrees, and the
integration times per sky pixel are 1200s, 600s and 120s at 24, 70 and
160~$\mu$m, respectively. The data were reduced and mosaicked using
the Data Analysis Tool \cite[]{gordon2005}.  We make use of a 
recent new analysis of the calibration by the instrument team and the
instrument support team that will soon be adopted officially. 
The uncertainty is now 4\%, 7\% and 12\% at 24, 70 and 160~$\mu$m,
respectively, and the calibration level has been changed by less than 10\%
compared with the previous determinations (See the Spitzer Science Center calibration
pages\footnote{http://ssc.spitzer.caltech.edu/mips/calib/conversion.html}). 

\cite{papovich2004} showed that the 80\% completeness level
at 24~$\mu$m in the GTO deep fields is reached at $S_{24} = 80 \mu$Jy.
Nevertheless, \cite{papovich2004} and \cite{chary2004} show that the
detection of very faint 24~$\mu$m sources, down to $S_{24} \sim 30
\mu$Jy, is possible, but with a increased photometric uncertainties and 
reduced completeness (to lower than 5\% at the GTO depth and 20\% at the GOODS
depth).

\cite{dole2004a} showed that at 70 and 160~$\mu$m sources can be
safely extracted down to $15$ mJy and $50$ mJy, respectively. The
\cite{frayer2005} results go deeper. However, confusion limits the
extraction of sources fainter than typically 56~$\mu$Jy at 24~$\mu$m,
3.2 mJy at 70~$\mu$m, and 36 mJy at 160~$\mu$m \cite[]{dole2004b}. A
priori information on the existence of a source deduced from shorter
wavelength and less confusion-limited observations can extend the
reliable detection threshold below this nominal confusion limit.

To implement this approach, we build a sample as follows:

\noindent $\bullet$ We select the central part of each field where all 3
MIPS wavelengths have a common sky coverage and maximum
redundancy. This area covers 0.29 Sq. Deg in the CDFS and LH, and 0.27
Sq. Deg. in HDFN, for a total of 0.85 Sq. Deg. in these three fields.

\noindent $\bullet$ In these selected areas, we identify every MIPS 24~$\mu$m 
source with $S_{24} \ge 60 \mu$Jy. This flux density limit corresponds
to 50\% completeness \cite[]{papovich2004}. There are 6543 galaxies above 
this limit in
CDFS, 6039 in the HDFN, and 6599 in the LH. The total number of
sources considered in the three fields is thus 19181.

To analyze this sample, we proceed as follows:

\noindent $\bullet$ In each field, we sort the 24~$\mu$m sources by
decreasing flux density $S_{24}$.

\noindent $\bullet$ We put the sources in 20 bins of flux density for
$S_{24} \ge 60 \mu$Jy. These bins have equal logarithmic width $\Delta
S_{24} / S_{24} \sim 0.15$, except for the bin corresponding to the
brightest flux, which includes all the bright sources (0.92mJy to
1Jy).

\noindent $\bullet$ We correct the average flux obtained by
stacking each $S_{24}$ bin for incompleteness, following the
correction of \cite{papovich2004} (their figure 1). Since the bins
between 60 and 80~$\mu$Jy are complete to the 50-80\% level, only the
weakest fluxes bins are significantly corrected.

%
\section{Stacking Analysis}
\label{section:StackingAnalysis}
The process of stacking the sources based on the 24~$\mu$m detections 
allows us to measure more of the total contribution
of 70 and 160~$\mu$m sources to the CIB.  

%
\subsection{Processing}
\label{section:Processing}

At 24~$\mu$m, the detector pixel size is 2.5 arcseconds, the FWHM 
of the point spread function (PSF) is 6
arcseconds, and the plate scale of the mosaic is chosen to be 1.25
arcseconds. At 70~$\mu$m, the detector pixel size is 9.9 arcseconds,
the FWHM of the PSF is 18 arcseconds, and the plate scale of the mosaic is chosen
to be 4.5 arcseconds. At 160~$\mu$m the detector pixel size is 18
arcseconds, the FWHM of the PSF is 40 arcseconds, and the plate scale of the
mosaic is chosen to be 18 arcseconds \cite[for more
details]{rieke2004,gordon2005}. The 70 and 160~$\mu$m mosaics have
been resampled to the scale of the 24~$\mu$m mosaic (1.25 arcseconds
per pixel) using a bilinear interpolation. This last step greatly
facilitates the weight management of the three maps, since each has 
different coverage, and it allows easy extraction of the signal at the three
wavelengths for the same sky position.

\begin{figure}[!t]
   \centering
   \includegraphics[width=0.4\textwidth]{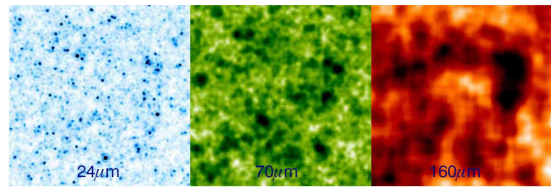}
   \caption{Images at 24, 70 and 160~$\mu$m (left to right) of stacked
   sources in the brightest bin of 24~$\mu m$ flux density, with a
   random position offset added before each sum. Color coding: dark is
   high flux, light is low flux. This allows us to check that the
   stacking method does not introduce any artifacts, i.e. a false
   source detection in the center. } \label{fig:imgoff}
\end{figure}

For each $S_{24}$ flux density bin we select every 24~$\mu$m source, 
extract a square image about 440 arcseconds on a side centered on
the source, and store it. We proceed similarly on the
mosaics at 70 and 160~$\mu$m, extracting images at the position of
each 24~$\mu$m source regardless of the presence of a detected FIR
source. The products at this stage are thus three cubes of data at 24,
70 and 160~$\mu$m with the same dimensions (same number of source images and
same box size) for each of the 24~$\mu$m flux density bins.

We then add the images in each cube at each wavelength, to generate a
stacked image of sources at 24, 70 and 160~$\mu$m for a given $S_{24}$
flux bin. This operation is a simple sum, without any outlier
rejection. When stacking, we rotate each image by $+\pi/2$ with
respect to the previous one (and so on), to cancel out the large-scale
background gradients such as the prominent zodiacal background at
24~$\mu$m.  This processing is done both in each field separately as
well as using all the data at once. Unless otherwise stated, we use
the stacked data of all the fields together in the rest of this paper.
We checked that no significant signal was detected when we added a
random or systematic artificial offset to each 24~$\mu$m position and
then performed exactly the same sub-image extraction and stacking as
we did for the real 24~$\mu$m source list. Figure~\ref{fig:imgoff}
shows the results of stacking the sources (only from the brightest
bin), with a random position offset added prior to the sum. No source
appears in the center, as expected. This guarantees that the stacking
method does not introduce an artifact that mimics a source.

Since the stacking analysis aims at statistically detecting faint
unresolved sources at 70 and 160~$\mu$m, in principle there is no need
to also stack data at 24~$\mu$m, where all sources are
resolved. However, doing so allows us to double-check the method,
since we know by design what the stacked photometry should be.

%
\begin{figure}[!t]
   \centering
   \includegraphics[width=0.4\textwidth]{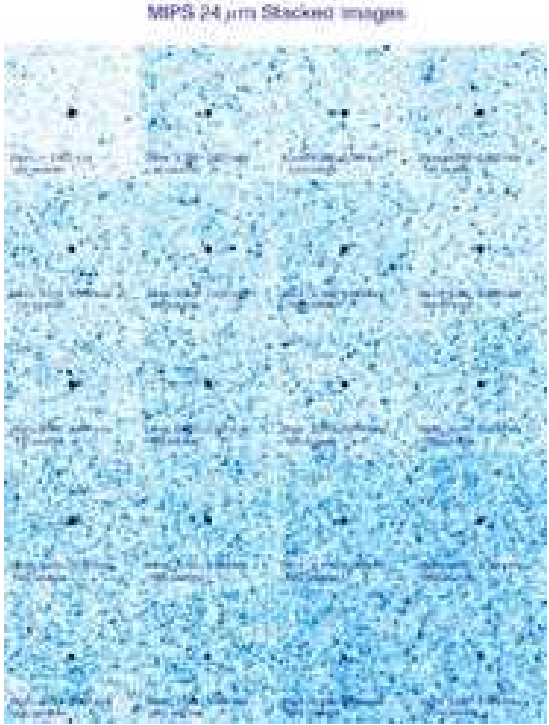}
   \caption{Images at 24~$\mu$m of stacked sources in bins of
   24~$\mu$m flux density ($S_{24} \ge 60\, \mu$Jy) in the three MIPS
   GTO Fields: CDFS, HDFN, and LH covering about 0.85 Sq. Deg. A total
   number of 19181 sources has been used. The number of sources used
   in the sum in each $S_{24}$ bin is reported. Each image has
   350 $\times$ 350 pixels of 1.25 arcsec, thus covering about 7.3
   $\times$ 7.3 Sq. Arcmin. Since no outlier rejection has been
   made, other sources can be seen in the surroundings of the stacked
   sources}. \label{fig:img024}
\end{figure}

\begin{figure}[!ht]
   \centering
   \includegraphics[width=0.4\textwidth]{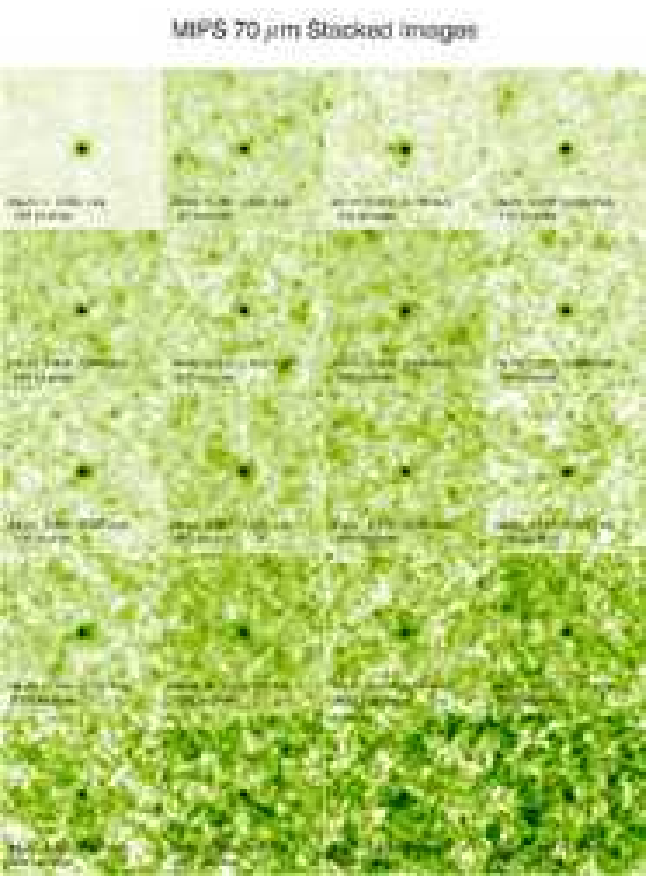}
   \caption{Images at 70~$\mu$m of stacked sources in bins of 24~$\mu
   m$ flux density. See caption of Figure~\ref{fig:img024} for details. } 
   \label{fig:img070}
\end{figure}

\begin{figure}[!hb]
   \centering
   \includegraphics[width=0.4\textwidth]{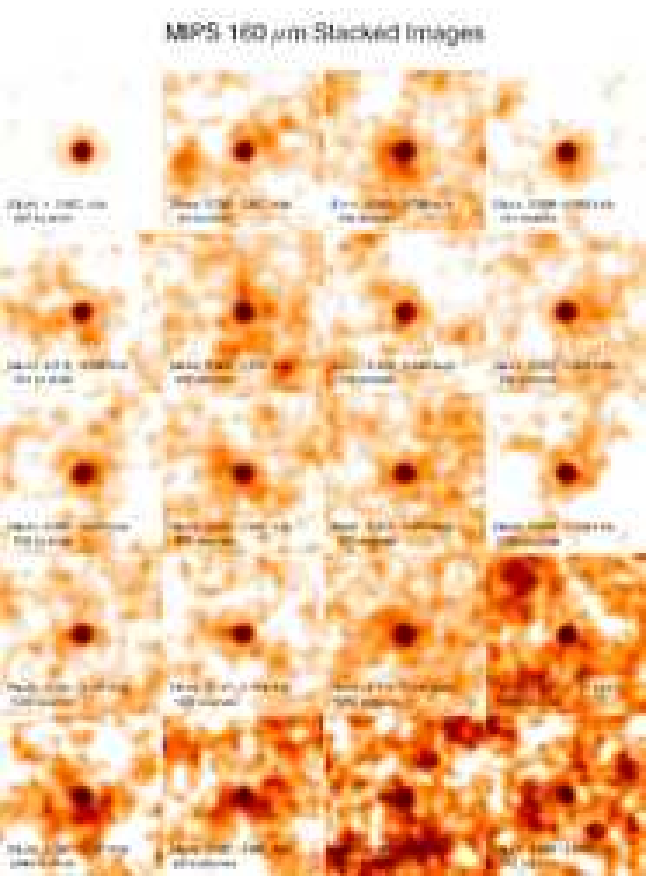}
   \caption{Images at 160~$\mu$m of stacked sources in bins of 24~$\mu
   m$ flux density. See caption of Figure~\ref{fig:img024} for details. } 
   \label{fig:img160}
\end{figure}

\begin{figure}[!t]
   \centering
   \includegraphics[width=0.5\textwidth]{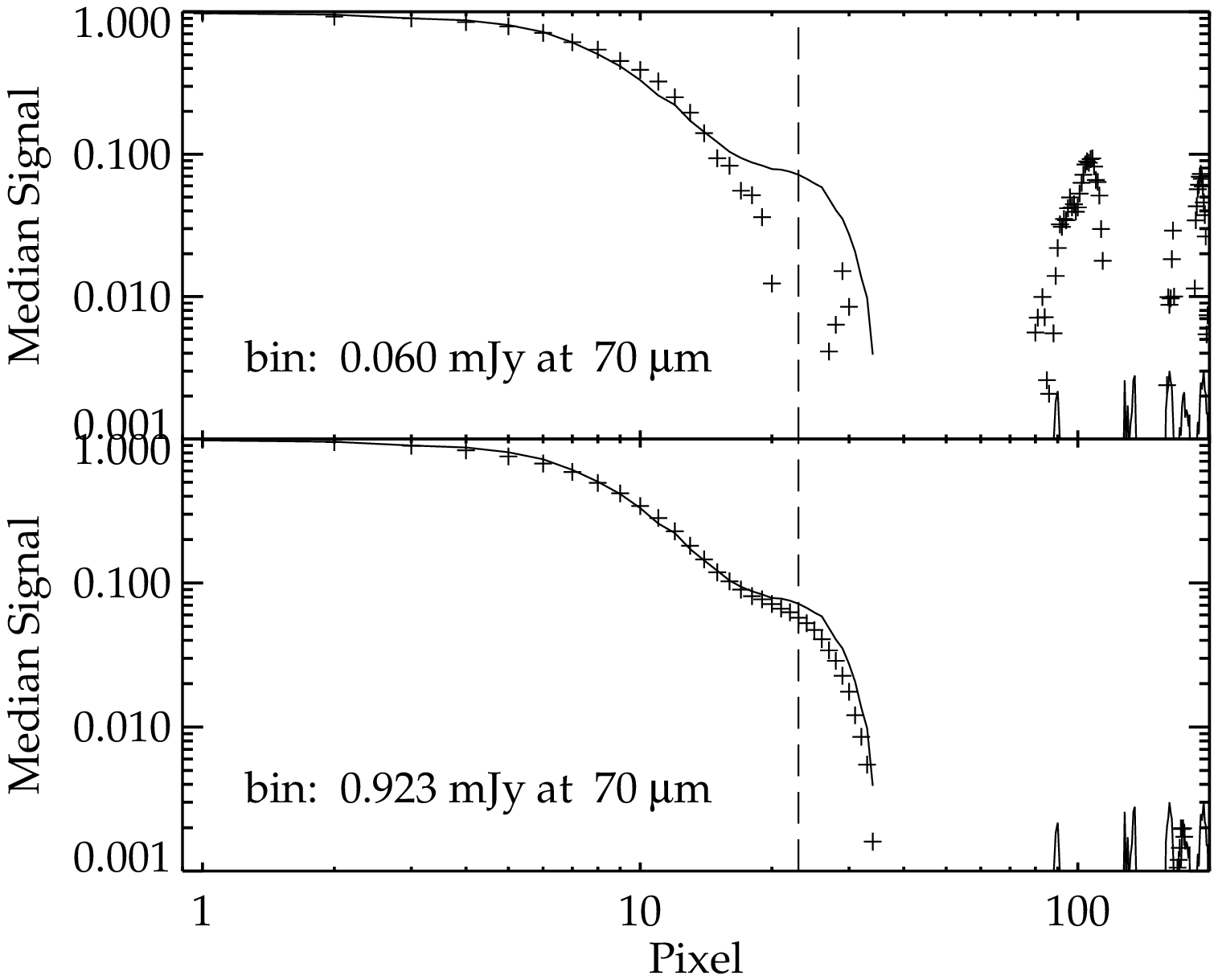}
   \includegraphics[width=0.5\textwidth]{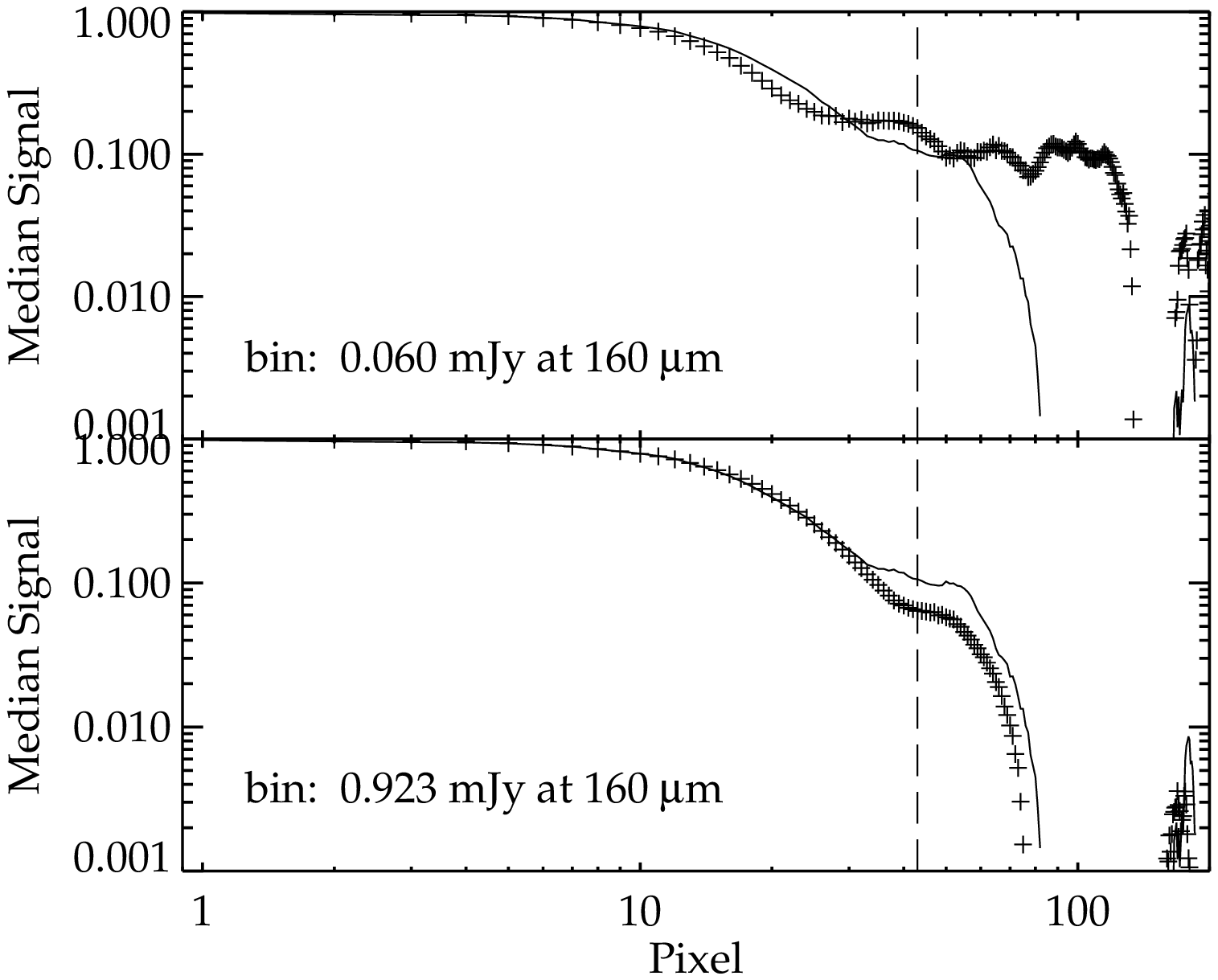}
   \caption{Normalized radial profiles of the stacked images. The
   crosses represent the data, and the solid line the empirical
   PSF. The vertical dotted lines show the radii of the aperture used
   for photometry. From top to bottom: faintest $S_{24}$ bin at
   70~$\mu$m; Brightest $S_{24}$ bin at 70~$\mu$m; Faintest $S_{24}$
   bin at 160~$\mu$m; Brightest $S_{24}$ bin at 160~$\mu$m .  } 
   \label{fig:radial_profiles}
\end{figure}

\begin{figure}[!t]
   \centering \includegraphics[width=0.5\textwidth]{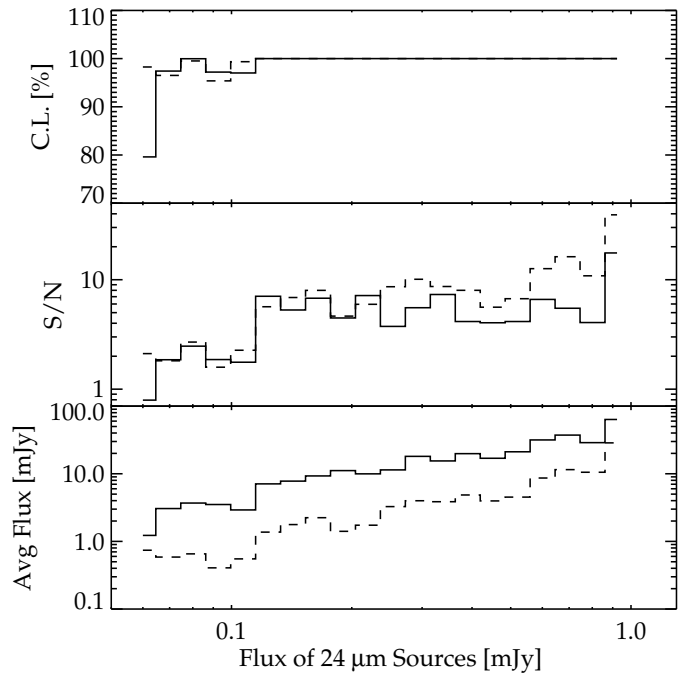}
   \caption{Top: Confidence Level of the detections at 70 (dash) and
   160~$\mu$m (solid) on stacked images (Fig.~\ref{fig:img070} and
   \ref{fig:img160}), as a function of the $S_{24}$ bin. Middle:
   Signal-to-Noise ratio, as computed from a Gaussian fit to the flux
   distribution measured on about 2000 positions; this S/N is not
   relevant at low flux (see Sect.~\ref{section:StackedImages}).
   Bottom: average flux in mJy per stacked source. Note that a
   different number of sources have been stacked in each bin.}
   \label{fig:sn_fir}
\end{figure}

%
\subsection{Stacked Images and Photometry}
\label{section:StackedImages}

The final stacked images at 24, 70 and 160~$\mu$m as a function of the
24~$\mu$m flux density $S_{24}$ are presented in
figures~\ref{fig:img024}, \ref{fig:img070}, and
\ref{fig:img160}, respectively. We report also in these figures the
number of sources stacked in each of the $S_{24}$ bins.  The figures
show clear detections of stacked sources at 70 and 160~$\mu$m for
every $S_{24}$ bin, even the faintest corresponding to $60 \le
S_{24} < 69 ~\mu $Jy. Given the surface density of the 24~$\mu$m
sources at 60~$\mu$Jy of $(9.6 \pm 0.04) \times 10^7$ sr$^{-1}$
\cite[]{papovich2004}, this translates to 1.04 and 0.2 beams per
source at respectively 70 and 160~$\mu$m \cite[using beams from their
Table 1]{dole2003}. This is well beyond the confusion limits at these
FIR wavelengths \cite[]{dole2004b}. This statistical detection of FIR
sources already demonstrates the great potential of this technique to
probe FIR galaxies down to levels below the confusion, thanks to
the excellent quality of the pointing and the stability of the
effective PSF (see below).

We check that the radial profile of the stacked sources is in
agreement with the PSF profile, at each wavelength and for each flux
bin. We show in Figure~\ref{fig:radial_profiles} two profiles at each
wavelength corresponding to the extreme cases: the brightest and
faintest $S_{24}$ flux bins. We used both the empirical PSF (from
bright sources) and the modeled STinyTim MIPS PSF
\cite[]{krist93,rieke2004,gordon2005}. At large $S_{24}$, the stacked
radial profiles at 70 and 160~$\mu$m (bottom plots in
Figure~\ref{fig:radial_profiles}) agree well with the PSF in the
central part. At the faintest fluxes (top plots), the agreement is
good down to about 10\% of the peak brightness. Since the stacked
images visually represent the 2-dimensional correlation function of
galaxies, the potential presence of many neighboring sources at small
scales (source clustering) might have widened the radial profile,
which is not observed; thus source clustering does not contribute
significantly to the noise budget.

We measure the flux density of the stacked sources with aperture
photometry and correct for aperture size. The radii of the
apertures and reference annulus are, in arcseconds: ($r_{aper}$,
$r_{int}$, $r_{ext}$) = (12.2, 17, 24), (30, 49, 79) and (54, 90, 126)
for 24, 70 and 160~$\mu$m, respectively. These radii correspond to
approximatively 3, 5 and 7 times the FWHM in the FIR and 2, 3, 4 times
the FWHM in the MIR.  We measured the noise in each image by using
about 2000 measurements on random positions. We compute the confidence
level (C.L.) of each detection (top of Figure~\ref{fig:sn_fir}) using
the cumulative distribution of the noise measurements. The deviation from
100\% of the C.L. is the probability that the noise creates a spurious
source.  For the faintest bin, the C.L. is around 80\%, and it rises to
97\% for the next four bins, and stays at 100\% for the brighter
$S_{24}$ bins. We fitted a Gaussian function to the distribution of 
noise to get the standard
deviation in order to estimate the S/N ratio. This method works for the brighter bins (middle panel in
Figure~\ref{fig:sn_fir}), where the flux distribution is indeed nearly
a Gaussian distribution. In this range, the S/N values have a median of 8 at
70~$\mu$m and 7 at 160~$\mu$m. In the three faintest bins, the
noise distribution is not Gaussian, because of the presence of
slightly brighter sources; the Gaussian fit is therefore not relevant and
we opt for the C.L. technique. The bottom plot in Figure 7 shows the average FIR
flux per stacked galaxy. A set of $\sim$100$\mu$Jy MIR-selected
galaxies would have a typical average FIR flux of $\sim$ 0.5 and
3 mJy if taken individually at 70 and 160~$\mu$m, respectively. Since
the confusion limits are at about 3 and 40 mJy at these wavelengths
\cite[]{dole2004b}, the gain of the stacking analysis technique is one
order of magnitude in flux compared to individual detection. 
Finally, it is not necessary to remove the brightest sources for the goals of this paper,
because we stack typically 1000 to 2000 galaxies per flux bin, so their
influence is negligible except maybe in the 3 faintest bins. 

\begin{figure}[!t]
   \centering
   \includegraphics[width=0.5\textwidth]{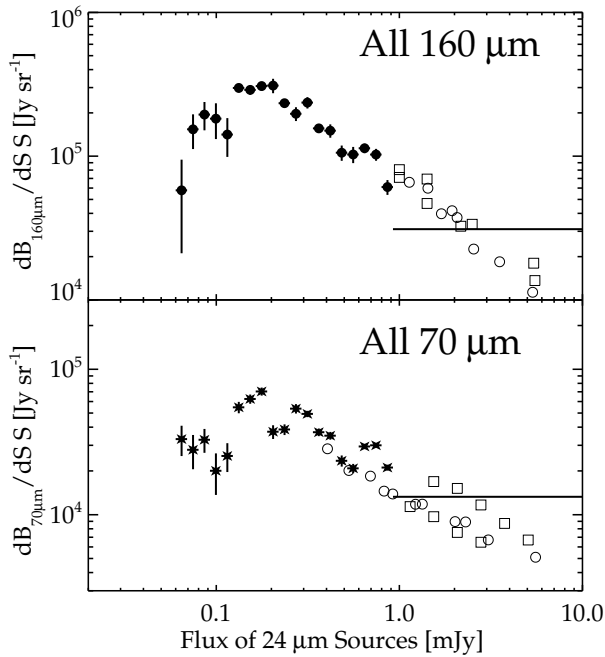}
   \caption{Contributions to the CIB: brightness of stacked sources at
   160 and 70~$\mu$m per logarithmic bin
   $\frac{dB_{\lambda}}{dlog(S_{24})}$ (solid symbols), as a function
   of the 24~$\mu$m flux, in all three MIPS GTO Fields. A completeness
   correction has been applied. The highest flux bin goes up to
   1Jy. Open symbols: published differential source counts multiplied
   by $S_{24}^{-0.5}$ and a color ratio of bright galaxies
   (160/24=60 and 70/24=20, cf the brightest bin in the lower
   panel of Fig.~{\ref{fig:sn_fir}}); Square: \cite{dole2004a};
   Circle: \cite{frayer2005}.  There is a good agreement between the
   source counts, the brightest stacked bin, and the fainter stacked
   bins. } \label{fig:bkg_diff}
\end{figure}

The brightness of the stacked sources at 70 and 160 per logarithmic
flux density bin, or $\frac{dB_{\lambda}}{dlog(S)}$, as a function of
the 24~$\mu$m flux bin, is presented in
figure~\ref{fig:bkg_diff}. $B_{\lambda}$ in MJy/sr is defined as the
total stacked flux density divided by the survey area. 
Using a logarithmic flux density bin allows
direct comparison of the contribution in energy of each bin to the
CIB, and is directly related to the differential source counts with a
scaling factor $S_{\nu}^{-0.5}$. In the range $100 \le S_{24} \le 300
\, \mu$Jy, both contributions to the CIB present a maximum, which
shows that the contributions have reached convergence.
Converting $S_{70}$ and $S_{160}$ into
$S_{24}$ using the color ratios of 9 and 30 (see 
Table~\ref{tab:CIBcolors} below), this means the FIR CIB will be mainly
resolved at $S_{70} \sim 0.1 \times 9 = 0.9$~mJy and at $S_{160} \sim
0.1 \times 30 = 3$~mJy. We have also plotted the source counts of
\cite{dole2004a} and \cite{frayer2005}; we used the conversion to
$S_{24}$ as given by color ratios relevant for bright galaxies of 20
and 60 at 70 and 160~$\mu$m, measured on the very bright
end of the bottom plot in Figure~\ref{fig:sn_fir}. Despite this
simplifying assumption of a single color ratio, there is excellent
agreement between the brightness derived from the stacking analysis
and the source counts. This plot can be used to constrain models of
galaxy evolution.

Sample variance plays a role in these results. To probe its effects,
we split each of our three fields (CDFS, HDFN, LH) into four subfields
of about 250 square arcmin each, and performed an independent analysis
on each of these twelve subfields. We obtain contributions varying in
some cases by as much as a factor of two (peak-to-peak).  For
instance, computing the standard deviation of the distribution of the
cumulative 160~$\mu$m flux (the faintest points in
Figure~\ref{fig:bkg}) measured over these 12 subfields gives
$\sigma=0.3 $ MJy/sr and a mean and median both of 0.53
MJy/sr. Renormalizing by the twelve sub-fields gives $\sigma=0.09$:
the uncertainty induced by the Large Scale Structure variations across
the fields is of order 15\%.


From here on in this paper, our error budget takes into account: 1) the
calibration uncertainties; 2) the photometric uncertainty; 3) the
large-scale structure (sample variance).

%
\section{Contributions of Mid-Infrared Galaxies to the Cosmic Infrared Background}
\label{section:ContribCIB}

%
\subsection{Value of the Cosmic Infrared Background Brightness}
\label{section:ValueCIB}

To compute the fraction of background resolved with the stacking
analysis of the MIPS data, we first need to review the measurements of
the total CIB, in particular at 24, 70 and 160~$\mu$m. It should be
remembered that the total cosmic background contains the contribution
of all extragalactic sources but also more diffuse emissions,
e.g. from dust in galaxy clusters \cite[]{montier2005}.  Furthermore
the extragalactic sources are expected to be mostly galaxies but it
cannot be excluded that other lower luminosity sources, population III
stars for instance, contribute significantly but will not be detected
directly in the present deep surveys. \\

\begin{figure*}[!t]
   \centering \includegraphics[width=0.90\textwidth]{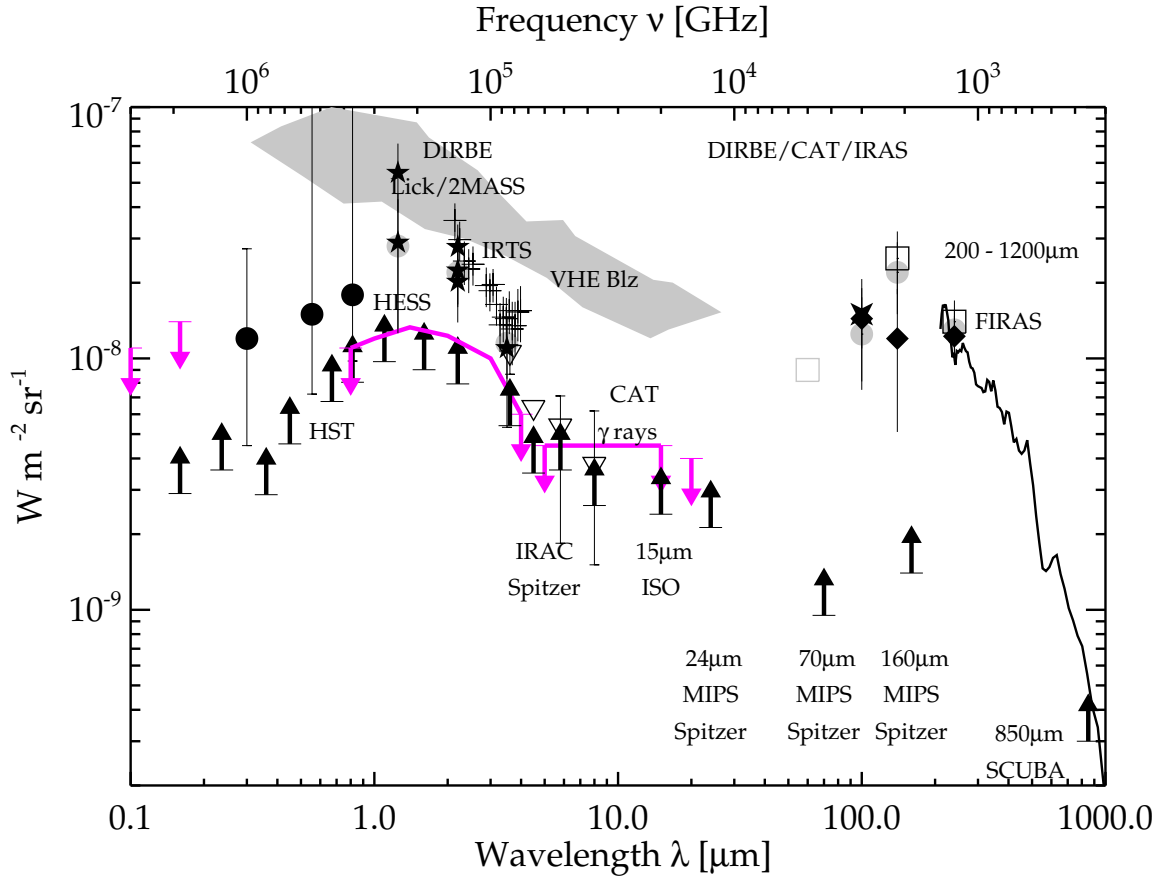}
   \caption{ Current measurements of the Extragalactic Background
   Light Spectral Energy Distribution from 0.1~$\mu$m to 1~mm, showing
   the Cosmic Optical Background (COB, with $\lambda \le 8~\mu$m) and
   the Cosmic Infrared Background (CIB, with $\lambda >
   8~\mu$m). Black arrows represent lower limits. Purple arrows and
   lines represent upper limits. The EBL observational constraints
   come from: \cite{edelstein2000} at 0.1~$\mu$m using Voyager UVS;
   \cite{brown2000} and \cite{gardner2000} with HST/STIS [lower
   limits]; \cite{madau2000} with HST (incl. NICMOS) and
   \cite{thompson2003}; \cite{bernstein2002} corrected by
   \cite{mattila2003} [filled circles]; \cite{matsumoto2005} between
   2.2 and 4~$\mu$m using the {\it IRTS} [thin plus];
   \cite{gorjian2000} at 2.2 and 3.3~$\mu$m using {\it DIRBE} and {\it
   Lick}; \cite{wright2001} and \cite{cambresy2001} at 1.25 and
   2.2~$\mu$m using {\it DIRBE} and {\it 2MASS} [five branch star];
   {\it DIRBE} values from \cite{wright2004} from 1.25 to 240~$\mu$m
   [gray circles]; {\it Spitzer IRAC} 3.6, 4.5, 5.8 and 8.0~$\mu$m
   lower limits from number counts by \cite{fazio2004a}; fluctuation
   analysis with IRAC by \cite{savage2005} (open triangles);
   \cite{schroedter2005} using Very High Energy Blazars, 98\%
   confidence upper limit [gray region]; H.E.S.S upper limit from
   \cite{aharonian2005} using P0.55 [solid line between 0.8 and
   4~$\mu$m]; \cite{renault2001} upper limits from 5 to 15~$\mu$m
   using the {\it CAT} in the $\gamma$-rays on Mkn501; \cite{elbaz99}
   lower limit at 15~$\mu$m using galaxy counts with {\it ISOCAM};
   upper limit at 20~$\mu$m by \cite{stecker97} on Mkn421; lower limit
   from galaxy counts at 24~$\mu$m with MIPS by \cite{papovich2004};
   an indirect evaluation at 60~$\mu$m using fluctuations in IRAS data
   from \cite{miville-deschenes2002} [open gray square]; lower limits
   at 70 and 160~$\mu$m using galaxy counts with MIPS by
   \cite{dole2004a}; an estimate of the CIB at 100~$\mu$m using {\it
   CAT} and {\it DIRBE} \cite{renault2001} [four branch star];
   \cite{lagache2000} at 100, 140 and 240~$\mu$m using {\it DIRBE} and
   WHAM, updated in the present work [diamond]; \cite{hauser98} at 140
   and 240~$\mu$m using {\it DIRBE} [open square]; \cite{smail2002}
   lower limit at 850~$\mu$m using galaxy counts with {\it SCUBA};
   \cite{lagache2000} spectrum between 200~$\mu$m and 1.2mm using {\it
   FIRAS} [solid line above 200~$\mu$m]. The IDL script to generate
   this figure is available on the web:
   http://www.ias.u-psud.fr/irgalaxies.  } \label{fig:cib_sed}
\end{figure*}

Measuring the CIB directly by photometry is particularly difficult
because one needs 1) an absolute photometer and 2) a proper estimate
of the foreground. The two FIR channels of MIPS are not absolute
photometers for the very extended spatial scales, since no internal
calibrated reference can be observed to calibrate absolutely the slow
response. A better knowledge of the instrument in the future may allow
a proper absolute calibration, using its ``Total Power'' mode,
similarly to what has been done successfully in the past, e.g. with
ISOPHOT at 170~$\mu$m on some selected low surface brightness fields
\cite[]{lagache2001}. However, MIPS is by design well calibrated for
small spatial scales, e.g., point sources, because the frequent
stimulator flashes and the scanning strategy (acting like a chopping
mode) properly track the fast response of the photoconductors
\cite[for instance]{gordon2005,gordon2006}. For these reasons, 
we do not use MIPS as an absolute photometer to estimate the level of
the CIB and the foregrounds, but as a detector of small scale
fluctuations to resolve the CIB based on 24~$\mu$m source
observations.  Our approach has the important advantage that it is not
biased by the foregrounds and their modeling, which can lead to
significant errors.

We therefore start by reviewing the direct measurements, using absolute
photometry in large beams, provided mainly by the COBE FIRAS and DIRBE
experiments and also the IRTS and rocket experiments in the near
infrared ($< 3\mu$m). These measurements can be combined with indirect upper
limits derived from observations of gamma rays from distant Blazars at
TeV energies.

To use the FIRAS and DIRBE data to provide CIB absolute measurements 
requires an accurate component separation. Local
extended emission from interplanetary and interstellar dust can be
removed using their specific SEDs and anisotropic spatial
distributions traced independently, as well as time variability for the
zodiacal emission and scattering \cite[for instance]{hauser2001}.
Early gamma ray data from Blazars from the CAT experiment led to
upper limits on the CIB intensity significantly lower than the DIRBE
residuals as pointed out by \cite{renault2001} and
\cite{wright2004}. Recent results on more distant Blazars
\cite[]{schroedter2005,aharonian2005} constrain the
CIB even more in the near and thermal infrared.  Together with lower limits
obtained by integrating the galaxy counts from {\it HST}, {\it ISO}, and {\it
Spitzer}, these measurements tightly constrain the Extragalactic
Background Light between $\sim 0.8$ to $\sim 20~\mu$m.

At 160~$\mu$m, the CIB can be interpolated from the DIRBE/COBE
measurements at 100~$\mu$m \cite[]{lagache2000} and 140 and 240~$\mu$m
\cite[]{hauser98}: 0.78$\pm$0.21, 1.17$\pm$0.32, 1.09$\pm$0.20
MJy/sr, respectively.  If the FIRAS photometric scale is used in the
calibration (rather than the DIRBE photometric calibration), lower
values are obtained at 140 and 240~$\mu$m: 0.7 MJy/sr and 1.02 MJy/sr
\cite[]{hauser98}. A large uncertainty in the determinations at 100
and 140~$\mu$m comes from the zodiacal emission removal, as is also
true at 60~$\mu$m.  The DIRBE zodiacal emission model was obtained by
\cite{kelsall98} relying on the variability with viewing geometry.
Its accuracy can be estimated {\it a posteriori} using the residuals
observed at wavelengths where the zodiacal emission is at a maximum (12
and 25~$\mu$m). The residual emission, obtained by \cite{hauser98},
has in fact a spectrum very similar to the zodiacal one.  The
residuals are about $4.7 \times 10^{-7}$ Wm$^{-2}$sr$^{-1}$ at 12 and
25~$\mu$m, far above the upper limit derived by high-energy
experiments like H.E.S.S. \cite[]{aharonian2005}, but not very much
larger than the uncertainties of the \cite{kelsall98} zodiacal
emission model. A conservative estimate of the amount of zodiacal
emission not removed in this model at 12 and 25~$\mu$m is therefore about
$4 \times 10^{-7}$ Wm$^{-2}$sr$^{-1}$. Using the \cite{kelsall98}
smooth high latitude zodiacal cloud colors, the amount not removed at
100, 140 and 240~$\mu$m translates to 0.30, 0.14, 0.045 MJy/sr,
respectively.  This reduces the CIB from 0.78 to 0.48 MJy/sr at
100~$\mu$m, from 1.17 to 1.03 at 140~$\mu$m and from 1.09 to 1.05 at
240~$\mu$m. Adopting the FIRAS photometric scale gives at 140 and
240~$\mu$m, 0.56 and 0.98 MJy/sr respectively.  From the above
discussion, we see that the CIB at 140 ~$\mu$m -- the closest in
wavelength to the 160~$\mu$m MIPS bandpass -- is still uncertain
by a factor of about 2 because of the uncertainty in the zodiacal
level. The DIRBE/FIRAS measurement of the CIB at 240~$\mu$m suffers
less from zodiacal residuals and photometric calibration uncertainty.

A firm upper limit of 0.3 MJy/sr at 60~$\mu$m has been derived by
\cite{dwek2005} using observations of TeV gamma ray emission from
distant AGNs. \cite{miville-deschenes2002} uses a fluctuation analysis
of IRAS maps to set an upper limit of 0.27 MJy/sr and give an estimate
of 0.18 MJy/sr, on the assumption that the level of
fluctuations-to-total intensity ratio is not strongly wavelength
dependent.

At 24~$\mu$m we use for the contribution of IR galaxies to the CIB
the estimate of \cite{papovich2004} of $2.7^{+1.1}_{-0.7}$
nW\,m$^{-2}$\,sr$^{-1}$. This value comes from 1) integration of the
source counts down to 60~$\mu$Jy giving $1.9 \pm 0.6$
nW\,m$^{-2}$\,sr$^{-1}$; 2) extrapolation of the source counts to
lower fluxes, giving a contribution of $0.8^{+0.9}_{-0.4}$
nW\,m$^{-2}$\,sr$^{-1}$; and 3) upper limits from \cite{stecker97} and
from CAT \cite[]{renault2001}.

The most constraining measurements and lower and upper limits on the
Cosmic Optical Background (COB) and the CIB from 0.1~$\mu$m to 1~mm
are all reported in Figure~\ref{fig:cib_sed}. The \cite{lagache2004}
model predicts a CIB at 240~$\mu$m of 0.98 MJy/sr, which is in very
good agreement with the estimate from combined measurements discussed
above. Furthermore this model agrees with the observational
constraints (e.g. number counts, CIB intensity and fluctuations). We can
thus take the CIB values from this model as a reasonable interpolation
between the better constrained CIB values at shorter and longer
wavelengths: 0.82 MJy/sr at 160~$\mu$m, and 0.15 MJy/sr at 70~$\mu$m.

\begin{table}[!t]
   \caption[]{Contribution to the CIB of $S_{24} \ge 60~\mu$Jy
galaxies. Sources of uncertainty come from photometry, calibration,
and large scale structure. The CIB column gives the best estimate
from the discussion in Sect.~\ref{section:ValueCIB}.}
   \label{tab:CIBcontrib}
   \begin{center}
   \begin{tabular}[h]{lcccc}
   \hline \\[-5pt]
        $\lambda$  & $\nu I_{\nu}$  & B$_{\lambda}$  &    CIB    & \% CIB \\
        $\mu$m     &  nW m$^{-2}$ sr$^{-1}$ & MJy/sr &   MJy/sr  & resolved\\[+5pt]
        \hline\\[-5pt]
                24      & $ 2.16 \pm  0.34 $  &  $  0.017 \pm 15 $\%  &  $  0.022 $  &  $   79 $ \\
                70      & $ 5.93 \pm  1.02 $  &  $  0.138 \pm 17 $\%  &  $  0.15 $  &  $   92 $ \\
                160     & $10.70 \pm  2.28 $  &  $  0.571 \pm 21 $\%  &  $  0.82 $  &  $   69 $ \\
    \hline \\
    \end{tabular}
        \begin{tabular}[h]{l}
        \end{tabular}
    \end{center}
\end{table}
%
%

%
\subsection{Contributions from MIR Sources with $S_{24} \ge 60 \mu$Jy}
\label{section:MIRSources}

\begin{figure}[!t]
   \centering
   \includegraphics[width=0.5\textwidth]{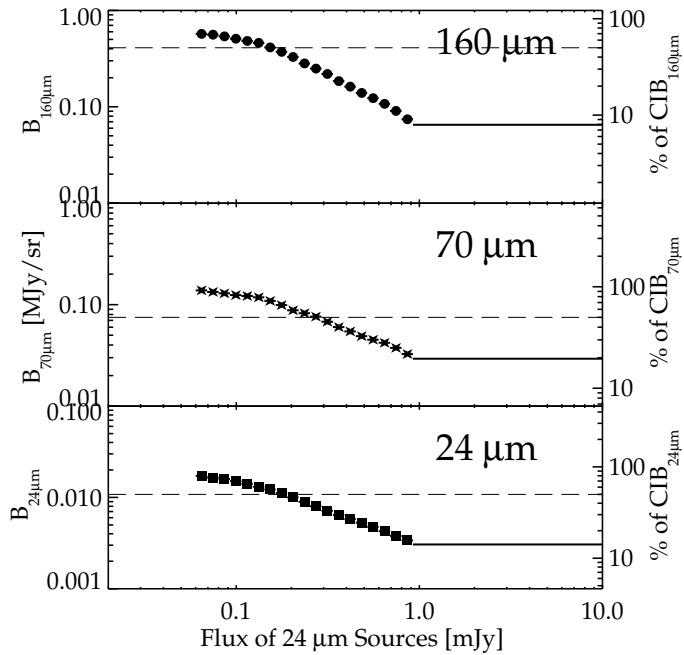}
   \caption{Cumulative brightness (left axis, in MJy/sr) and fraction
   of the background resolved (right axis) at 160, 70 and 24~$\mu$m
   (from top to bottom), as a function of the 24~$\mu$m flux in the three combined MIPS
   GTO Fields. A completeness correction has been applied based on the
   incompleteness level of the 24~$\mu$m sample. The horizontal dashed
   line represents 50\% of the CIB at each wavelength. } 
   \label{fig:bkg}
\end{figure}

To estimate the contribution of MIR sources to the background, we
add up the brightnesses of all the $S_{24}$ bins to get the integrated
light at 24, 70 and 160 of all the resolved 24~$\mu$m sources. Each
$S_{24}$ bin is corrected for incompleteness.  The results are
presented Table~\ref{tab:CIBcontrib}, and Figure~\ref{fig:bkg} shows
the cumulative integrated light from galaxies in the FIR as a function
of $S_{24}$. For a sanity check, we obtain that at 24 microns the percentage of 
the CIB that is resolved is 79\%, which is in agreement with
\cite{papovich2004} within the error bars. At 70 and 160 micron we
resolve 92\% and 69\% of the background, respectively.

Half of the 24~$\mu$m CIB is resolved by sources with $S_{24} \ge 190
\mu$Jy. In the FIR, half of the 70~$\mu$m CIB is resolved by 24~$\mu$m
sources brighter than $S_{24} \sim 220 \mu$Jy, and half of the
160~$\mu$m CIB is resolved by 24~$\mu$m sources brighter than $S_{24}
\sim 130 \mu$Jy. This difference between 70 and 160 
suggests that the CIB at 160~$\mu$m is dominated by galaxies at
slightly higher redshift than at 70~$\mu$m, a consequence of the
spectral shape of LIRGs and ULIRGs or, equivalently, the effect of
k-correction. This point is illustrated in figures 5 and 6 of
the review by \cite{lagache2005}.

\begin{figure*}[!t]
   \centering
   \includegraphics[width=1.0\textwidth]{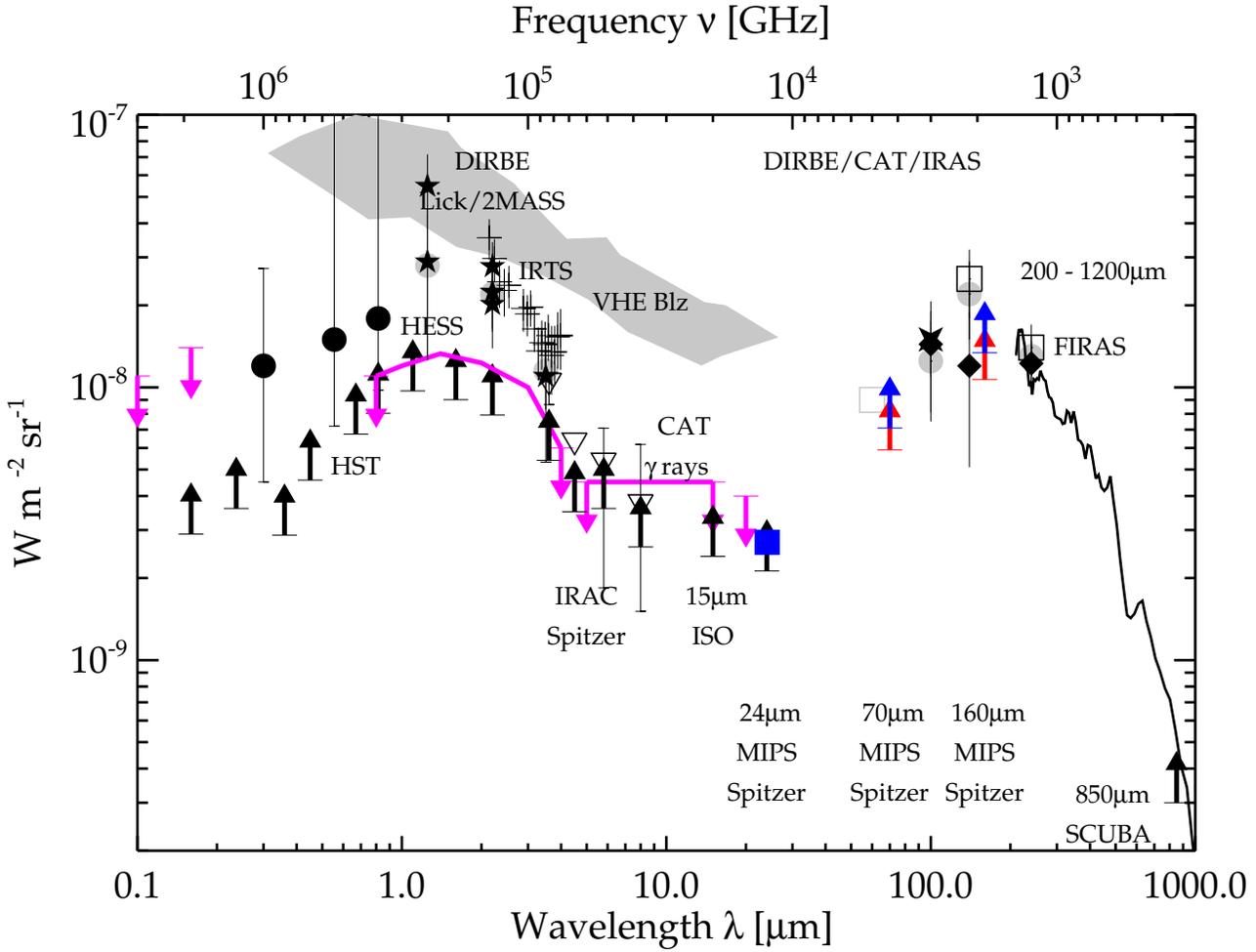}
   \caption{Extragalactic Background Light Spectral Energy
   Distribution from 0.1~$\mu$m to 1~mm, with new constraints from
   MIPS. Red arrows (lower limits) represent the fraction of the CIB
   resolved at 70 and 160~$\mu$m using the stacking analysis for
   sources with $S_{24} \ge 60~\mu$Jy. The blue square represents the
   contribution of all the unresolved 24~$\mu$m sources (extrapolation
   from number counts), and the blue arrows represent the contribution
   of all 24~$\mu$m sources to the FIR background, using a simple
   color extrapolation for 70 and 160~$\mu$m
   (Sect.~\ref{section:NewLowerLimits}). See Figure~\ref{fig:cib_sed}
   for the other symbols.}  \label{fig:cib_sed_stack}
\end{figure*}

To put in perspective the problem of resolving the CIB and what the
stacking analysis accomplishes, we plot in
figure~\ref{fig:cib_sed_stack} the new observed constraints on the
extragalactic background SED.  The fraction of the CIB resolved at
MIPS wavelengths by unbiased surveys was 79\%, 20\% and 7\% at
respectively 24, 70 and 160~$\mu$m \cite[]{papovich2004,dole2004a}.
When using the present stacking analysis, this fraction rises to 92\%
and 69\% at 70 and 160~$\mu$m respectively, and is represented by the
red lower limits (see also Table~\ref{tab:CIBcontrib}).

Based purely on observations without modeling of galaxy SEDs, we 
find that most of the FIR background is
resolved into MIR galaxies. This confirms the model-dependent result of
\cite{elbaz2002}.  This analysis is the first direct resolution of the
CIB simultaneously in the MIR and the FIR.

Moreover, we can now securely establish the physical parameters of the
typical galaxies responsible for most of the CIB near its peak.
Previous studies based on ISO already characterized the 15~$\mu$m
population
\cite[for instance]{flores99,elbaz2003,franceschini2003}; 
see \cite{lagache2005} for a review. Our 24~$\mu$m sample is almost
complete in flux (80\% completeness down to $S_{24} = 80 \mu$Jy and
50\% at $60 \mu$Jy), and the physical properties of $S_{24} \ge 80
\mu$Jy galaxies have been extensively studied 
\cite[]{lefloch2004,perez-gonzalez2005,lefloch2005,caputi2005a}. These
works, mainly targeting the CDFS field, show that 25-30\% of the
24~$\mu$m galaxies lie at redshifts $z \ge 1.5$, and that the redshift
distribution peaks around $z \sim 1$ (between 0.7 and 1.1). Assuming
the CDFS is a representative field, the MIR and FIR CIB is thus mainly
composed of galaxies with typical redshifts of unity, with a
contribution from $z >1.5$ galaxies. At these redshifts, the galaxies
are mostly LIRGs with typical bolometric luminosities of about $3
\times 10^{11}\, L_{\odot}$ (between $10^{11}$ and $10^{12} \,
L_{\odot}$) forming about 50 $M_{\odot} yr^{-1}$ (20-130). They have
intermediate stellar masses of about $10^{10}$ to $10^{11} \,
M_{\odot}$ \cite[]{perez-gonzalez2005,caputi2005a}. From this latter
work we can also estimate the specific star formation rates of
these galaxies to be between 0.1 and 1 Gyr$^{-1}$.

%
\subsection{Mean colors of the galaxies contributing to the CIB}
\label{section:ColorsContrib}

Looking at the 24~$\mu$m number counts of \cite{papovich2004}, one can
see that the bulk of 24~$\mu$m CIB is mainly due to sources with $130
\le S_{24} \le 400~\mu$Jy. We select three cuts in $S_{24}$ to
investigate the colors of the contributions to the CIB by different
galaxy populations. In the following, redshifts come from
\cite{caputi2005a} (see their figure 5), and the relative
contributions come from the integration of the \cite{papovich2004}
source counts and the \cite{lagache2004} model. The cuts are:

\noindent $\bullet$  Above 400~$\mu$Jy: bright galaxies 
contributing about 25\% to the 24~$\mu$m CIB. The redshift
distribution has a mean of 0.53 and a median of 0.44.

\noindent $\bullet$  $130 \le S_{24} \le 400~\mu$Jy: galaxies
contributing the most to the 24~$\mu$m CIB, about 30\%. The
redshift distribution has a mean of 1.18 and a median of 1.03.

\noindent $\bullet$  $60 \le S_{24} \le 130~\mu$Jy: fainter galaxies
with relatively low contributions to the 24~$\mu$m CIB (about
15\%). The redshift distribution has a mean of 1.27 and a median of
1.11. The sample becomes incomplete at $60 \le S_{24} \le 80
\mu$Jy, so the mean redshift may be underestimated. 

The observed colors change systematically with $S_{24}$, as can
be seen in the bottom of figure~\ref{fig:sn_fir}: $S_{70}$ vs
$S_{24}$ shows a slope larger than one, when $S_{160}$ vs $S_{24}$
shows an average slope of the order of
unity. Table~\ref{tab:CIBcolors} gives the colors and their associated
$1\sigma$ uncertainties in the three bins.  The 160/24 color is
compatible with a constant of 33. The 70/24 color increases with the
flux from 6.3 to 9.6. Finally, the 160/70 color steadily decreases with
flux.

\begin{table}[!t]
   \caption[]{Mean observed colors in $I_{\nu}$ of $S_{24} \ge
    60~\mu$Jy galaxies contributing to the CIB.}
   \label{tab:CIBcolors}
   \begin{center}
   \begin{tabular}[h]{lccc}
   \hline \\[-5pt]
  $S_{24}$ in $\mu$Jy & 160/70 & 160 / 24 & 70 / 24 \\[+5pt]
  \hline\\[-5pt]
  $S_{24} \ge 400$      & $   3.2 \pm    0.4 $ & $ 29.7 \pm    3.8$ & $   9.6 \pm    0.8 $ \\
  $130 \le S_{24} \le 400$ & $   4.4 \pm    0.5 $ & $ 40.6 \pm    4.2$ & $   9.4 \pm    0.9 $ \\
  $60 \le S_{24} \le 130$ & $   5.3 \pm    1.6 $ & $ 32.7 \pm    6.8$ & $   6.3 \pm    1.1 $ \\[+5pt]
    \hline \\[-5pt]
  CIB$^a$       &   5.5  &  38.0 &   6.9 \\[+5pt]
    \hline \\
    \end{tabular}
        \begin{tabular}[h]{l}
            $^a$ Data and Model; See Sect.~\ref{section:ValueCIB}\\
        \end{tabular}
    \end{center}
\end{table}
%

These colors can be interpreted as the SED of a LIRG being redshifted,
since fainter 24~$\mu$m sources lie at larger redshifts: the 160/70
ratio increases (with decreasing flux) because the peak of the big grains' 
FIR spectrum
is shifted longwards of 160~$\mu$m. The color ratios involving the
24~$\mu$m band are less obvious to interpret, since the Polycyclic
Aromatic Hydrocarbon (PAH) \cite[]{puget89} features (especially
between 6.2 and 8.6~$\mu$m) and the silicate absorption feature are
redshifted into and then out of this band. The 70/24 ratio evolution
might have for its origin a mix of PAH (increasing the 24) and
very small grains continuum (decreasing the 70) being redshifted, that
cancel out each other.

If one wants to extrapolate the contribution of fainter ($S_{24} \le
60~\mu$Jy) MIR galaxies to the FIR CIB, a conservative approach is to
use a constant 160/24 and 70/24 color ratio for the unresolved
population. To set these ratios, we take the colors from the 
faintest population ($60 \le S_{24} \le 130~\mu$Jy); this faint
population presumably has the closest characteristics to the
unresolved one.  We will therefore use $I_{160}/I_{24} = 32.7 \pm 6.8$, and
$I_{70}/I_{24} = 6.3 \pm 1.1$ (from Table~\ref{tab:CIBcolors}). Since
the contribution to the CIB of these faint galaxies is modest
(30\% at most), the large uncertainties in these color ratios will
not dominate the total background estimate.

%
\section{New Estimates of the Cosmic Far-Infrared Background}
\label{section:NewLowerLimits}

%
\subsection{New Lower Limits at 70 and 160~$\mu$m}

The present stacking analysis performed on detected galaxies $S_{24}
\ge 60 \, \mu$Jy gives strong measured lower limits to the CIB
due to galaxies at 70 and 160~$\mu$m, without requiring any
modeling. To determine upper limits to the FIR CIB requires
a different approach. There are many difficulties at 70~$\mu$m in
extracting an accurate value of the CIB, mostly due to the
problems in the removal of the zodiacal component 
\cite[for instance]{finkbeiner2000,renault2001}. At 160~$\mu$m the CIB
estimate is more robust, but still with a significant uncertainty
(factor of $\sim$3, see Sect.~\ref{section:ValueCIB}).

Another way to get a good estimate of the FIR galaxy CIB brightness is to
estimate the unresolved 24~$\mu$m background fraction, use the 160/24
and 70/24 colors measured for the weakest sources, and then apply
these colors to the unresolved part to get the 70 and 160~$\mu$m
background estimates. Thus, we extrapolate
the colors of galaxies with $S_{24} \le 60\,\mu$Jy using the 
colors of the $60 \le S_{24} \le 130\,\mu$Jy galaxies 
derived in the previous section. To estimate
the unresolved 24~$\mu$m background, \cite{papovich2004} used a simple
extrapolation of the differential number counts. Since the slope of
the counts below 100~$\mu$Jy is strongly decreasing ($-1.5 \pm 0.1$ in
$dN/dS$), the integral is dominated by the largest fluxes
$S_{24}$. The estimate is robust, unless a hypothetical faint
population exists. The remaining unresolved 24~$\mu$m background
created by $S_{24} < 60,\mu$Jy sources is therefore 0.54 $nW m^{-2}
sr^{-1}$, (to be compared to 2.16 $nW m^{-2} sr^{-1}$ for
$S_{24}>60,\mu$Jy sources).

We derive the extrapolated FIR CIB level due to IR galaxies using:
\begin{equation}
\nu I_{\nu}(\lambda) = \nu I_{\nu}(24) \times
\frac{I_{\lambda}}{I_{24}} \times \frac{24}{\lambda}
\end{equation}

The results of the extrapolation are presented in
Table~\ref{tab:CIBextrapolation}. We obtain $7.1 \pm 1.0$ and $ 13.4
\pm 1.7$~$nW m^{-2} sr^{-1}$, at 70 and 160~$\mu$m respectively.  Our
new estimate, based on the integration of all the 24~$\mu$m IR
galaxies, is in principle a lower limit because it does not account
for any diffuse emission unrelated to the IR galaxies, nor for a small
fraction of IR galaxies that might have been missed. Indeed, the
extrapolation in color of the unresolved 24~$\mu$m population accounts
for the faint-end of the luminosity function, but not for the
hypothetical very high-redshift sources, or faint local galaxies
with high FIR output, like a hypothetical population of elliptical galaxies with 
large 160/24 colors. However, if this population exists, its
contribution to the FIR background is constrained by the upper limits
to be less than $\sim$20\%.

Our estimate at 70~$\mu$m is higher than the \cite{lagache2004} model
estimate by 11\%, and lower by about 13\% at 160~$\mu$m.
About 25\% of the CIB brightness at 70 and 160~$\mu$m comes from faint
MIR sources ($S_{24} \le 60~\mu$Jy).  Assuming our new FIR CIB values
represent the actual CIB values, we estimate that our stacking
analysis of $S_{24} \ge 60\,\mu$Jy galaxies finally resolves 75-80\%
of the background at 70 and 160~$\mu$m. We also show that the
population dominating the CIB is made of galaxies seen at 24~$\mu$m
and their simplest extrapolation to lower fluxes.

\begin{table}[!t]
   \caption[]{Contributions of the 24$~\mu$m galaxies to the FIR CIB
                     in nW m$^{-2}$ sr$^{-1}$. For the $S _{24}\le 60~\mu$Jy galaxies,
                     a simple color extrapolation has been used, as described in
                     Sect.~\ref{section:NewLowerLimits}.}
   \label{tab:CIBextrapolation}
   \begin{center}
   \begin{tabular}[h]{lccc}
   \hline \\[-5pt]
          &     24~$\mu$m      &        70~$\mu$m       &  160~$\mu$m \\[+5pt]
        \hline\\[-5pt]
        $> 60\,\mu$Jy  & $  2.16 \pm  0.26 $ & $  5.9 \pm   0.9$ & $ 10.7 \pm   1.6 $ \\
        $< 60\,\mu$Jy$^a$ & $  0.54 $ & $  1.2 \pm   0.2$ & $  2.6 \pm   0.5 $ \\
        {\bf total CIB$^b$} &  2.7$^c$ & {\bf $  7.1 \pm   1.0$} & {\bf $ 13.4 \pm   1.7$} \\
        CIB prior $^c$  &  2.7$^c$  &  6.4$^c$  & 15.4$^c$ \\[+5pt]
    \hline \\
    \end{tabular}
        \begin{tabular}[h]{l}
        $^a$ Estimate using an extrapolation from 60 to 0 $\mu$Jy.\\
        $^b$ CIB estimate due to IR galaxies.\\
        $^c$ Data and Model; See discussion Sect.~\ref{section:ValueCIB}\\
        \end{tabular}
    \end{center}
\end{table}
%
%

\begin{figure}[!t]
   \centering
   \includegraphics[width=0.50\textwidth]{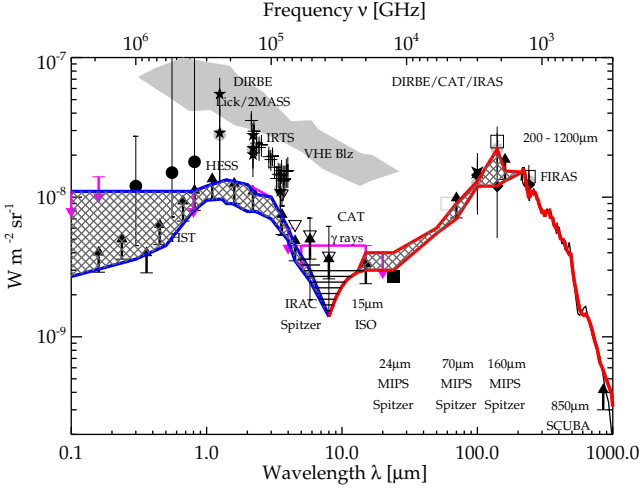}
   \caption{Cosmic Optical Background and Cosmic Infrared Background
   due to galaxies permitted zone estimate (shaded area), using upper
   and lower values. See Figure~\ref{fig:cib_sed} for the other
   symbols. } \label{fig:cib_sed_eblzone}
\end{figure}

%
\subsection{Spectral Energy Distribution of the Extragalactic Background}

In the near and mid-IR, upper and lower limits tightly constrain the
EBL SED: 1) with {\it HST}+{\it Spitzer} and H.E.S.S between 0.8 and
4~$\mu$m, and 2) with {\it ISO}, {\it Spitzer} and CAT between 5 and
24~$\mu$m. In this range, the EBL SED is constrained to better than
50\% (and to the 20\% level in several wavelength ranges). The EBL is
now also well constrained in the FIR; direct measurements of the
diffuse emission and our new lower limits constrain the CIB SED to the
50\% level.

The permitted zone for the EBL SED is presented in
Figure~\ref{fig:cib_sed_eblzone}. This zone is defined as the area
between current upper and lower limits. In this zone, the COB
brightness ranges from 19.5 to 35.5~nW m$^{-2}$ sr$^{-1}$, and the CIB
from 24 to 27.5~nW m$^{-2}$ sr$^{-1}$. The ratio COB/CIB thus ranges
from 0.7 to 1.5.

From these constraints, we may derive a conservative estimate of the
EBL SED, that typically lies between the upper and lower limits and
that makes use of well known physical processes. The CIB estimate,
based on the \cite{lagache2004} model, agrees with the data and is
strongly constrained in the MIR and the 240-400~$\mu$m range. It
strongly decreases with increasing frequency below 8~$\mu$m because of
the main PAH features at 6.2 to 8.6~$\mu$m being redshifted. The COB
estimate also decreases with increasing wavelength above 2~$\mu$m
because of the old stellar population SED. This simple SED behavior is
in agreement with the model of \cite{primack99}. Our reasonable guess
is that the COB and CIB have equal contributions around 8~$\mu$m.

\begin{figure}[!t]
   \centering
   \includegraphics[width=0.50\textwidth]{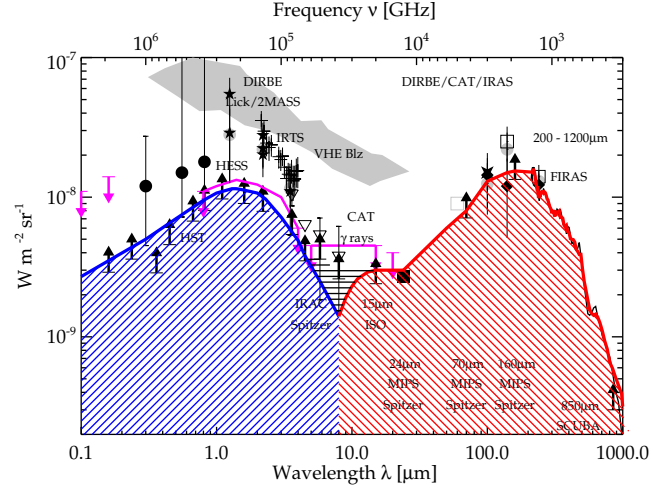}
   \caption{Our best Cosmic Optical Background (blue-shaded) and
   Cosmic Infrared Background (red-shaded) estimates. The gray-shaded
   area represents the region of overlap. See Figure~\ref{fig:cib_sed}
   for the other symbols. }  \label{fig:cib_sed_ebl}
\end{figure}

Figure~\ref{fig:cib_sed_ebl} shows our smooth EBL SED estimate (thick
line), as well as our best estimate of the COB (blue shaded) and the
CIB (red shaded). The overlap region where both COB and CIB contribute
significantly and the resulting total EBL is shown as the gray-shaded
area around 8~$\mu$m.  We find that the brightness of the COB is 23 nW
m$^{-2}$ sr$^{-1}$, and 24 nW m$^{-2}$ sr$^{-1}$ for the CIB. The
ratio between the COB and CIB is thus of the order of unity for this
EBL SED.

Our results are in contradiction with \cite{wright2004} who finds a
COB/CIB ratio of 1.7, and values at least 50\% higher than ours: 59 nW
m$^{-2}$ sr$^{-1}$ (COB) and 34 nW m$^{-2}$ sr$^{-1}$ (CIB).  However,
the \cite{wright2004} estimate came before the strong upper limits of
H.E.S.S \cite[]{aharonian2005} below 4~$\mu$m. This limit puts the COB
much closer to the integrated light from galaxy counts than to the
diffuse measurements. From the galaxy counts and stacking analysis
(lower limits), and high-energy experiments (upper limits), the EBL is
now very well constrained. In particular, we can now securely state
that the contributions to the EBL of faint diffuse emissions outside
identified galaxy populations -- too weak to be detected in current
surveys, like population III stars relic emission, galaxy clusters,
hypothetical faint IR galaxy populations -- can represent only a small
fraction of the integrated energy output in the universe.

%
\subsection{The Extragalactic Background vs the Cosmic Microwave Background}

\begin{figure}[!t]
   \centering
   \includegraphics[width=0.50\textwidth]{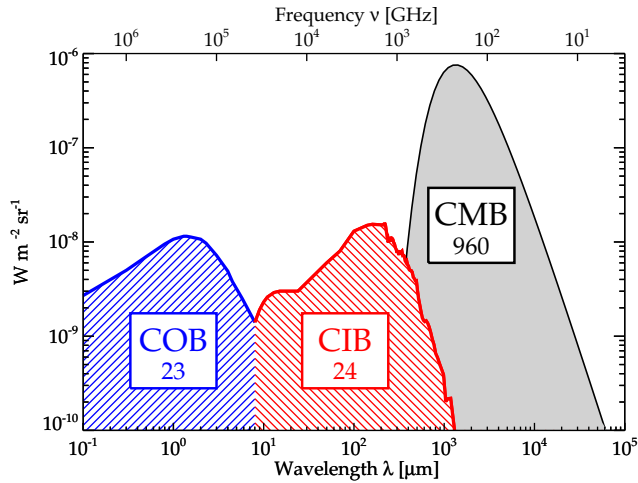}
   \caption{Schematic Spectral Energy Distributions of the most
   important (by intensity) backgrounds in the universe, and their
   approximate brightness in nW m$^{-2}$ sr$^{-1}$ written in the
   boxes. From right to left: the Cosmic Microwave Background (CMB),
   the Cosmic Infrared Background (CIB) and the Cosmic Optical
   Background (COB).  } \label{fig:cib_cmb}
\end{figure}

It is interesting to update the contributions of the most
intensive electromagnetic backgrounds in the universe, as has been
done for instance by \cite{scott2000} or \cite{wright2004}, and we
schematically represent these in Figure~\ref{fig:cib_cmb}. Obviously,
the Cosmic Microwave Background (CMB) dominates the universe's SED,
and accounts for about 960 nW m$^{-2}$ sr$^{-1}$.  We showed that the
CIB and COB each account for 23 and 24 nW m$^{-2}$ sr$^{-1}$,
respectively.  With a total of 47 nW m$^{-2}$ sr$^{-1}$ in the optical
and the Far-Infrared, the EBL represents about 5\% of the brightness
of the CMB.  Taking into account the complete SED of the EBL will not
change this picture, since the contributions to the total EBL
brightness of the radio, UV, X-ray \cite[]{mushotzky2000,
hasinger2001} and $\gamma$ ray \cite[]{strong2004} extragalactic
backgrounds are smaller by one to three orders of magnitude than the
COB and CIB \cite[]{scott2000}.

The galaxy formation and evolution processes provide 5\% in
brightness of the electromagnetic content of the Universe. Half of
the energy comes in the form of starlight (COB) and half as
dust-reprocessed starlight (CIB).  The maximum of the power
distribution is at $\sim 1.3~\mu$m for the COB and $\sim 150~\mu$m for
the CIB (Fig.~\ref{fig:cib_cmb}).  There are therefore on average 115
infrared photons for 1 visible photon emitted in these processes.

%
\section{Conclusions}
\label{section:Conclusion}

Our key points and results for the resolution and characterization
of the FIR CIB and the EBL are:

\noindent $\bullet$ A stacking analysis in three fields covering 0.85
square degrees including a sample of 19181 MIPS 24~$\mu$m sources with $S_{24} \ge
60\,\mu$Jy lets us probe faint 70 and 160~$\mu$m galaxies one order of 
magnitude below
the confusion level and with a high
signal-to-noise ratio. We take into account in our noise budget
uncertainties coming from: photometry, calibration systematics, and
large-scale structure.

\noindent $\bullet$ 24~$\mu$m galaxies down to $S_{24} = 60 \, \mu$Jy
contribute 79\%, 92\%, 69\% of the CIB at respectively 24, 70 and
160~$\mu$m (using 2.7, 6.4 and 15.4~nW m$^{-2}$ sr$^{-1}$ as the total CIB
values at 24, 70 and 160~$\mu$m, respectively). This is
the first direct measurement of the contribution of MIR-selected
galaxies to the FIR background.

\noindent $\bullet$ We derive the contributions to the CIB by flux density
bin, and show good agreement between our stacking analysis and the
published source counts. This is a strong constraint for models.
Moreover, we show that the CIB will be mainly resolved at flux
densities of about $S_{70} \sim 0.9$~mJy and $S_{160} \sim 3$~mJy
at 70 and 160~$\mu$m, respectively.

\noindent $\bullet$ We directly measure that the total CIB, peaking
near 150~$\mu$m, is largely resolved into MIR galaxies. Other works
\cite[especially]{perez-gonzalez2005,lefloch2005,caputi2005a} show
that these MIPS 24~$\mu$m sources are $\sim 3 \times 10^{11} \,
L_{\odot}$ LIRGs distributed at redshifts $z \sim 1$, with stellar
masses of about $3 \times 10^{10}$ to $3 \times 10^{11} \, M_{\odot}$
and specific star formation rates in the range 0.1 to 1 Gyr$^{-1}$.

\noindent $\bullet$ Using constant color ratios 160/24 and 70/24 for
MIR galaxies fainter than 60~$\mu$Jy, we derive new conservative lower
limits to the CIB at 70 and 160~$\mu$m including the faint IR galaxies
undetected at 24~$\mu$m: $7.1 \pm 1.0$ and $13.4 \pm 1.7$~nW~m$^{-2}$
~sr$^{-1}$, respectively. These new estimates agree within 13\% with the
\cite{lagache2004} model.

\noindent $\bullet$ Using these new estimates for the 70
and 160$\mu$m CIB, we show that our stacking analysis down to $S_{24} \ge 60
\, \mu$Jy resolves $>$75\% of the 70 and 160~$\mu$m CIB.

\noindent $\bullet$ Upper limits from high-energy experiments and
direct detections together with lower limits from galaxy counts and
stacking analysis give strong constraints on the EBL SED.

\noindent $\bullet$ We estimate the Extragalactic Background Light (EBL)
Spectral Energy Distribution (SED) permitted zone (between lower and
upper limits), and measure the optical background (COB) to be in the
range 19.5-35.5~nW~m$^{-2}$~sr$^{-1}$, and the IR background (CIB) in
the range 24 to 27.5~nW~m$^{-2}$~sr$^{-1}$. The ratio COB/CIB thus
lies between 0.7 and 1.5.

\noindent $\bullet$ We integrate our best estimate of the COB and the
CIB, and obtain respectively 23 and 24 nW m$^{-2}$ sr$^{-1}$; We
find a COB/CIB ratio close to unity.

\noindent $\bullet$ The galaxy formation and evolution processes have
produced photons equivalent in brightness to 5\% of the CMB, with
equal amounts from direct starlight (COB) and from dust-reprocessed
starlight (CIB). We compute that the EBL produces on average 115
infrared photons per visible photon.

\begin{acknowledgements}
This work is based on observations made with the {\it Spitzer}
Observatory, which is operated by the Jet Propulsion Laboratory,
California Institute of Technology under NASA contract 1407.  We thank
the funding from the MIPS project, which is supported by NASA through
the Jet Propulsion Laboratory, subcontract \#1255094. This work also
benefited from funding from the CNES (Centre National d'Etudes
Spatiales) and the PNC (Programme National de Cosmologie). We thank
Jim Cadien for the great help in the data processing. We thank Nabila
Aghanim, Herv\'e Aussel, Noel Coron, Daniel Eisenstein, David Elbaz, 
Charles Engelbracht, Dave Frayer, Karl Gordon, Nicolas Ponthieu, Martin
Schroedter, and Xianzhong Zheng for fruitful discussions.
\end{acknowledgements}

\begin{thebibliography}{  }

\bibitem[{Aharonian} et~al.(2005)]{aharonian2005}
{Aharonian}, F, {Akhperjanian}, A.~G, {Bazer-Bachi}, A.~R, {Beilicke}, M, \&
  {H.E.S.S Collaboration}.
\newblock Nature, 2006, accepted, astro-ph/0508073.

\bibitem[{Bernstein} et~al.(2002)]{bernstein2002}
{Bernstein}, R.~A, {Freedman}, W.~L, \& {Madore}, B.~F.
\newblock 2002, {\em \ApJ}, 571:107.

\bibitem[{Brown} et~al.(2000)]{brown2000}
{Brown}, T.~M, {Kimble}, R.~A, {Ferguson}, H.~C, {Gardner}, J.~P, {Collins},
  N.~R, \& {Hill}, R.~S.
\newblock 2000, {\em \AJ}, 120:1153.

\bibitem[{Cambresy} et~al.(2001)]{cambresy2001}
{Cambresy}, L, {Reach}, W.~T, {Beichman}, C.~A, \& {Jarrett}, T.~H.
\newblock 2001, {\em \ApJ}, 555:563.

\bibitem[{Caputi} et~al.(2006)]{caputi2005a}
{Caputi}, K.~I, {Dole}, H, {Lagache}, G, et al., 
\newblock 2006, {\em \ApJ}, 637:727.

\bibitem[{Chary} \& {Elbaz}(2001)]{chary2001}
{Chary}, R \& {Elbaz}, D.
\newblock 2001, {\em \ApJ}, 556:562.

\bibitem[{Chary} et~al.(2004)]{chary2004}
{Chary}, R, {Casertano}, S, {Dickinson}, M.~E, et al., 
\newblock 2004, {\em \ApJS}, 154:80.

\bibitem[{Dole} et~al.(2003)]{dole2003}
{Dole}, H, {Lagache}, G, \& {Puget}, J.~L.
\newblock 2003, {\em \ApJ}, 585:617.

\bibitem[{Dole} et~al.(2004a)]{dole2004a}
{Dole}, H, {Le Floc'h}, E, {Perez-Gonzalez}, P.~G, et al., 
\newblock 2004a, {\em \ApJS}, 154:87.

\bibitem[{Dole} et~al.(2004b)]{dole2004b}
{Dole}, H, {Rieke}, G.~H, {Lagache}, G, et al.,
\newblock 2004b, {\em \ApJS}, 154:93.

\bibitem[{Dole}(2003)]{dole2003a}
{Dole}, H.
\newblock In {Gry}, C, {Peschke}, S.~B, {Matagne}, J, et al. editors,
{\em Exploiting the ISO Data Archive},  page 307. ESA SP-511, 2003.
astro-ph/0211310

\bibitem[{Dwek} \& {Krennrich}(2005)]{dwek2005}
{Dwek}, E \& {Krennrich}, F.
\newblock 2005, {\em \ApJ}, 618:657.

\bibitem[{Edelstein} et~al.(2000)]{edelstein2000}
{Edelstein}, J, {Bowyer}, S, \& {Lampton}, M.
\newblock 2000, {\em \ApJ}, 539:187.

\bibitem[{Egami} et~al.(2004)]{egami2004}
{Egami}, E, {Dole}, H, {Huang}, J.~S, et al., 
\newblock 2004, {\em \ApJS}, 154:130.

\bibitem[{Elbaz} \& {Cesarsky}(2003)]{elbaz2003}
{Elbaz}, D \& {Cesarsky}, C.~J.
\newblock 2003, {\em \Sci}, 300:270.

\bibitem[{Elbaz} et~al.(1999)]{elbaz99}
{Elbaz}, D, {Cesarsky}, C.~J, {Fadda}, D, et al., 
\newblock 1999, {\em \AaA}, 351:L37.

\bibitem[{Elbaz} et~al.(2002)]{elbaz2002}
{Elbaz}, D, {Flores}, H, {Chanial}, P, et al., 
\newblock 2002, {\em \AaA}, 381:L1.

\bibitem[{Elbaz}(2005)]{elbaz2005}
{Elbaz}, D.
\newblock in Space Science Reviews ISO Special Issue "ISO science
legacy - a compact review of ISO major achievements", Ed.C.Cesarky \&
A.Salama (Springer)
\newblock astro-ph/0503389, 2005.

\bibitem[{Fazio} et~al.(2004)]{fazio2004a}
{Fazio}, G.~G, {Ashby}, M. L.~N, {Barmby}, P, et al., 
\newblock 2004, {\em \ApJS}, 154:39.

\bibitem[{Finkbeiner} et~al.(2000)]{finkbeiner2000}
{Finkbeiner}, D.~P, {Davis}, M, \& {Schlegel}, D.~J.
\newblock 2000, {\em \ApJ}, 544:81.

\bibitem[{Flores} et~al.(1999)]{flores99}
{Flores}, H, {Hammer}, F, {D\'esert}, F.~X, et al., 
\newblock 1999, {\em \AaA}, 343:389.

\bibitem[{Franceschini} et~al.(2003)]{franceschini2003}
{Franceschini}, A, {Berta}, S, {Rigopoulou}, D, et al.,
\newblock 2003, {\em \AaA}, 403:501.

\bibitem[{Frayer} et~al.(2006)]{frayer2005}
{Frayer}, D.~T., {Fadda}, D., {Yan}, L., et al.,
\newblock 2006, {\em \AJ}, 131:250.

\bibitem[{Gardner} et~al.(2000)]{gardner2000}
{Gardner}, J.~P, {Brown}, T.~M, \& {Ferguson}, H.~C.
\newblock 2000, {\em \ApJ}, 542:L79.

\bibitem[{Genzel} \& {Cesarsky}(2000)]{genzel2000}
{Genzel}, R \& {Cesarsky}, C.~J.
\newblock 2000, {\em \ARAA}, 38:761.

\bibitem[{Gispert} et~al.(2000)]{gispert2000}
{Gispert}, R, {Lagache}, G, \& {Puget}, J.~L.
\newblock 2000, {\em \AaA}, 360:1.

\bibitem[{Gordon} et~al.(2005)]{gordon2005}
{Gordon}, K.~D, {Rieke}, G.~H, {Engelbracht}, C.~W, et al., 
\newblock 2005, {\em \PASP}, 117:503.

\bibitem[{Gordon} et~al.(2006)]{gordon2006}
{Gordon}, K.~D, {Bailin}, J., {Engelbracht}, C.~W, et al., 
\newblock 2005, {\em \ApJ}, in press, astro-ph/0601314

\bibitem[{Gorjian} et~al.(2000)]{gorjian2000}
{Gorjian}, V, {Wright}, E.~L, \& {Chary}, R.~R.
\newblock 2000, {\em \ApJ}, 536:550.

\bibitem[{Hasinger} et~al.(2001)]{hasinger2001}
{Hasinger} G., {Altieri} B., {Arnaud} M., et al.,
\newblock 2001, {\em \AaA}, 365, L45

\bibitem[{Hauser} \& {Dwek}(2001)]{hauser2001}
{Hauser}, M.~G \& {Dwek}, E.
\newblock 2001, {\em \ARAA}, 37:249.

\bibitem[{Hauser} et~al.(1998)]{hauser98}
{Hauser}, M.~G, {Arendt}, R.~G, {Kelsall}, T, et al., 
\newblock 1998, {\em \ApJ}, 508:25.

\bibitem[{Houck} et~al.(2005)]{houck2005}
{Houck}, J.~R, {Soifer}, B.~T, {Weedman}, D, et al., 
\newblock 2005, {\em \ApJ}, 622:L105.

\bibitem[{Kashlinsky}(2005)]{kashlinsky2005}
{Kashlinsky}, A.
\newblock 2005, {\em \PhR}, 409:361.

\bibitem[{Kelsall} et~al.(1998)]{kelsall98}
{Kelsall}, T, {Weiland}, J.~L, {Franz}, B.~A, et al., 
\newblock 1998, {\em \ApJ}, 508:44.

\bibitem[{Krist}(1993)]{krist93}
{Krist}, J.
\newblock Tiny Tim : an HST PSF simulator.
\newblock In {Hanisch}, R.~J, {Brissenden}, J.~V, \& {Barnes}, J, editors, {\em
  Astronomical Data Analysis Software and Systems II}, page 536. A.S.P.
  Conference Series, 1993.

\bibitem[{Lagache} et~al.(1999)]{lagache99}
{Lagache}, G, {Abergel}, A, {Boulanger}, F, et al., 
\newblock 1999, {\em \AaA}, 344:322.

\bibitem[{Lagache} et~al.(2000)]{lagache2000}
{Lagache}, G, {Haffner}, L.~M, {Reynolds}, R.~J, \& {Tufte}, S.~L.
\newblock 2000, {\em \AaA}, 354:247.

\bibitem[{Lagache} \& Dole(2001)]{lagache2001}
{Lagache}, G, \& {Dole}, H.
\newblock 2001, {\em \AaA}, 372:702

\bibitem[{Lagache} et~al.(2003)]{lagache2003}
{Lagache}, G, {Dole}, H, \& {Puget}, J.~L.
\newblock 2003, {\em \MNRAS}, 338:L555.

\bibitem[{Lagache} et~al.(2004)]{lagache2004}
{Lagache}, G, {Dole}, H, {Puget}, et al., 
\newblock 2004, {\em \ApJS}, 154:L112.

\bibitem[{Lagache} et~al.(2005)]{lagache2005}
{Lagache}, G, {Puget}, J.~L, \& {Dole}, H.
\newblock 2005, {\em ARAA}, 43, 727.

\bibitem[{Le Floc'h} et~al.(2004)]{lefloch2004}
{Le Floc'h}, E, {P\'erez-Gonz\'alez}, P.~G, {Rieke}, et al., 
\newblock 2004, {\em \ApJS}, 154:L170.

\bibitem[{Le Floc'h} et~al.(2005)]{lefloch2005}
{Le Floc'H}, E, {Papovich}, C, {Dole}, H, et al., 
\newblock 2005, {\em \ApJ}, 632:L169.

\bibitem[{Lonsdale} et~al.(2004)]{lonsdale2004}
{Lonsdale}, C, {Polletta}, M. M.~C, {Surace}, J, et al., 
\newblock 2004, {\em \ApJS}, 154:L54.

\bibitem[{Madau} \& {Pozzetti}(2000)]{madau2000}
{Madau}, P \& {Pozzetti}, L.
\newblock 2000, {\em \MNRAS}, 312:L9.

\bibitem[{Matsumoto} et~al.(2005)]{matsumoto2005}
{Matsumoto}, T, {Matsuura}, S, {Murakami}, H,  et al., 
\newblock 2005, {\em \ApJ}, 626:31.

\bibitem[{Mattila}(2003)]{mattila2003}
{Mattila}, K.
\newblock 2003, {\em \ApJ}, 591:119.

\bibitem[{Miville-Desch\^enes} et~al.(2002)]{miville-deschenes2002}
{Miville-Desch\^enes}, M.~A, {Lagache}, G, \& {Puget}, J.~L.
\newblock 2002, {\em \AaA}, 393:749.

\bibitem[{Montier} \& {Giard}(2005)]{montier2005}
{Montier}, L.~A \& {Giard}, M.
\newblock 2005, {\em \AaA}, 439:35.

\bibitem[{Mushotzky} et~al.(2000)]{mushotzky2000}
{Mushotzky}, R. F,. and {Cowie},  L. L., and {Barger}, A. J,. and {Arnaud}, K. A. 
\newblock 2000, Nature, 404:459

\bibitem[{Papovich} et~al.(2004)]{papovich2004}
{Papovich}, C, {Dole}, H, {Egami}, E, et al., 
\newblock 2004, {\em \ApJS}, 154:70.

\bibitem[{Partridge} \& {Peebles}(1967)]{partridge67}
{Partridge}, R.~B. \& {Peebles}, P.~J.~E. 
\newblock 1967, {\em \ApJ}, 148:377

\bibitem[{P\'erez-Gonz\'alez} et~al.(2005)]{perez-gonzalez2005}
{P\'erez-Gonz\'alez}, P.~G, {Rieke}, G.~H, {Egami}, E, et al., 
\newblock 2005, {\em \ApJ}, 630:82.

\bibitem[{Primack} et~al.(1999)]{primack99}
{Primack}, J.~R, {Bullock}, J.~S, {Somerville}, R.~S, \& {Macminn}, D.
\newblock 1999, {\em \APh}, 11:93.

\bibitem[{Puget} \& {Leger}(1989)]{puget89}
{Puget}, J.~L \& {Leger}, A.
\newblock 1989, {\em \ARAA}, 27:161.

\bibitem[{Puget} et~al.(1996)]{puget96}
{Puget}, J.~L, {Abergel}, A, {Bernard}, J.~P, et al., 
\newblock 1996, {\em \AaA}, 308:L5.

\bibitem[{Renault} et~al.(2001)]{renault2001}
{Renault}, C, {Barrau}, A, {Lagache}, G, \& {Puget}, J.~L.
\newblock 2001, {\em \AaA}, 371:771.

\bibitem[{Rieke} et~al.(2004)]{rieke2004}
{Rieke}, G.~H, {Young}, E.~T, {Engelbracht}, C.~W, et al., 
\newblock 2004, {\em \ApJS}, 154:25.

\bibitem[{Sanders} \& {Mirabel}(1996)]{sanders96}
{Sanders}, D.~B \& {Mirabel}, I.~F.
\newblock 1996, {\em \ARAA}, 34:749.

\bibitem[{Savage} \& {Oliver}(2005)]{savage2005}
{Savage}, R.~S \& {Oliver}, S.
\newblock 2005, {\em \MNRAS}, submitted, astro-ph/0511359

\bibitem[{Schroedter}(2005)]{schroedter2005}
{Schroedter}, M.
\newblock 2005, {\em \ApJ}, 628:617.

\bibitem[{Scott}(2000)]{scott2000}
{Scott}, D.
\newblock 2000, ``Cosmic Flows Workshop'', ASP Conference Series,
Stephane Courteau and Jeffrey Willick Eds, p. 403

\bibitem[{Smail} et~al.(2002)]{smail2002}
{Smail}, I, {Ivison}, R.~J, {Blain}, A.~W, \& {Kneib}, J.~P.
\newblock 2002, {\em \MNRAS}, 331:495.

\bibitem[{Soifer} \& {Neugebauer}(1991)]{soifer91}
{Soifer}, B.~T \& {Neugebauer}, G.
\newblock 1991, {\em \AJ}, 101:354.

\bibitem[{Stecker} \& {De Jager}(1997)]{stecker97}
{Stecker}, F.~W \& {De Jager}, O.~C.
\newblock 1997, {\em \ApJ}, 476:712.

\bibitem[{Strong} et~al.(2004)]{strong2004}
{Strong}, A. W. and {Moskalenko} I. V. and {Reimer} 0.
\newblock 2004, {\em \ApJ}, 613:956.

\bibitem[{Thompson} (2003)]{thompson2003}
{Thompson}, R.~I
\newblock 2003, {\em \ApJ}, 596:748.

\bibitem[{Werner} et~al.(2004)]{werner2004}
{Werner}, M.~W, {Roellig}, T.~L, {Low}, F.~J, et al., 
\newblock 2004, {\em \ApJS}, 154:1.

\bibitem[{Wright}(2001)]{wright2001}
{Wright}, E.~L.
\newblock 2001, {\em \ApJ}, 553:538.

\bibitem[{Wright}(2004)]{wright2004}
{Wright}, E.~L.
\newblock 2004, {\em New Astronomy Reviews}, 48:465.

\bibitem[{Xu} et~al.(2001)]{xu2001}
{Xu}, C, {Lonsdale}, C.~J, {Shupe}, D.~L, et al., 
\newblock 2001, {\em \ApJ}, 562:179.

\end{thebibliography}

\end{document}